\documentclass[prl,aps, superscriptaddress,twocolumn]{revtex4-1}
\usepackage{amsmath,amssymb}
\usepackage{graphicx}
\usepackage{wasysym}
\usepackage{amsfonts}
\usepackage{bm}
\usepackage{enumerate}
\usepackage{color}
\usepackage[resetlabels]{multibib}
\usepackage{epstopdf}
\usepackage{latexsym}
\usepackage[breaklinks,colorlinks = true,linkcolor = red,urlcolor=cyan,citecolor=red]{hyperref}
\usepackage[caption=false,singlelinecheck=false]{subfig}

\usepackage{times}
\newcommand{\bea}{\begin{eqnarray}}
\newcommand{\eea}{\end{eqnarray}}
\newcommand{\be}{\begin{eqnarray}}
\newcommand{\ee}{\end{eqnarray}}
\newcommand{\bw}{\begin{widetext}}
\newcommand{\ew}{\end{widetext}}

\begin{document}
\title{Magnetic field and thermal Hall effect in a pyrochlore U(1) quantum spin liquid}

\author{Hyeok-Jun Yang}
\email{yang267814@kaist.ac.kr}
\affiliation{Department of Physics, Korea Advanced Institute of Science and Technology, Daejeon, 34141, Korea}
\author{Hee Seung Kim}
\email{shamoo0829@kaist.ac.kr}
\affiliation{Department of Physics, Korea Advanced Institute of Science and Technology, Daejeon, 34141, Korea}
\author{SungBin Lee}
\email{sungbin@kaist.ac.kr}
\affiliation{Department of Physics, Korea Advanced Institute of Science and Technology, Daejeon, 34141, Korea}
\date{\today}
\begin{abstract}
The antiferromagnetic system on a rare-earth pyrochlore has been focused as a strong candidate of U(1) quantum spin liquid. Here, we study the phase transitions driven by external magnetic field and discuss the large thermal Hall effect due to emergent spinon excitations with staggered gauge fields. 
Despite the spinons, the charge excitations of the effective action that carry spin-1/2 quantum number, do not couple to the external field, 
the emergent U(1) gauge field is influenced in the presence of external magnetic field.
Especially along [111] and [110]-directions, we discuss the possible phase transitions between U(1) spin liquids with different gauge fluxes are stabilized in fields. Beyond the cases where gauge flux per plaquette are fixed to be either 0 or $\pi$, there exists a regime where the staggered gauge fluxes are stabilized without time reversal symmetry. In such a phase, the large thermal Hall conductivity $\kappa_{xy}/T\sim 4.6\times 10^{-3}\text{W}/(\text{K}^2\cdot\text{m})$ is expected to be observed below 1K.
\end{abstract}
\maketitle

\label{sec:Introduction}
{\textbf{\textit {  Introduction ---}}}
The exotic phases of matter have broaden the manners to understand the strongly correlated electron systems\cite{anderson1973resonating, anderson1987resonating, fradkin2013field}. Especially, the quantum spin liquid (QSL) whose long-ranged order is suppressed even at zero-temperature have demanded a new framework to understand the internal orders other than conventional ones\cite{balents2010spin, savary2016quantum, zhou2017quantum}. One of the fascinating peculiarity of the QSLs is the existence of non-local excitations resulted from the quantum entanglement. They characterize the nature of the low-energy excitations which carry fractional quantum numbers. Unfortunately, the experimental probe is suffered from the inevitable non-locality\cite{balents2010spin, lacroix2011introduction, wynn2001limits, savary2016quantum}. Nonetheless, several predictions which can be deduced from their low-energy excitations might be the indirect methodology to unveil the phenomena of QSLs. 

The rare earth pyrochlore materials with a chemical formula RE$_2$TM$_2$O$_7$ contain several
candidate materials to realize QSLs\cite{harris1997geometrical, canals1998pyrochlore, bramwell2001spin, nakatsuji2006metallic, molavian2007dynamically, gardner2010magnetic, thompson2011rods, ross2011quantum, chang2012higgs, kimura2013quantum, pan2014low, pan2016measure, gao2019experimental}. Microscopically, the interplay between highly localized nature of the $f$-electrons in rare-earth ions, strong spin-orbit coupling and the crystal field results in the effective pseudospin-$1/2$ model on the pyrochlore lattice\cite{onoda2010quantum, onoda2011quantum, onoda2011effective, huang2014quantum, princep2015crystal, ruminy2016crystal}. Thus, the pseudospin at each site $i$ is represented as $S_i^{\pm}, S_i^{z}$ about their local $\hat{z}$ axes which are towards the center of a tetrahedron. Due to the geometrical frustration, the magnetic moment is disordered even at extremely low temperature. In the absence of quantum fluctuations, the massively degenerate ground states so called ``two-in two-out" states are realized\cite{ramirez1999zero, moessner2001magnets, moessner2006geometrical, castelnovo2008magnetic}. In the presence of quantum fluctuations induced by $S_i^{\pm}$, however, it diagonalizes the ground state manifold to give rise to the fractionalized liquid phase with emergent U(1) gauge structure, dubbed U(1) QSL\cite{moessner2003three, hermele2004pyrochlore, sikora2009quantum, savary2012coulombic, lee2012generic, shannon2012quantum}.

In this paper, we study the U(1) QSLs subject to the external magnetic field uniformly. The U(1) QSLs evolve differently as the field direction changes, we consider two case, [110] and [111]-directions. There is no charge carriers inside the system, thus the degree of the freedom for the low-energy excitations do not couple to the external field through the Peierls substitution. However, the Zeeman coupling with the magnetic dipoles generates the nearest-neighbor spinon hopping. As long as the U(1) QSL persists, i.e. the external field is not too strong, the effective action standing for the emergent gauge field is modified. This is reflected in the permeability, eventually the photon speed\cite{lantagne2017electric}. Importantly, the spinon propagation is coupled to the emergent gauge flux, thus the dynamics of spinon can be controlled by the external field. The spinon band structure manifest them which are responsible for the physical observables.

Focusing on the rare-earth pyrochlores described by Kramers doublet for pseudospin (the system contains odd number of electrons in RE$^{3+}$), 
we perform the standard perturbation in magnetic field $\textbf{B}$ for the lowest order correction in the coupling constants\cite{hermele2004pyrochlore}. 
Considering on the spinon and the associated gauge field, we show there are phase transitions in the U(1) QSL with different fluxes\cite{wen2002quantum, lee2012generic, chen2017spectral}.
We draw the schematic phase diagrams and find the regime where a new type of flux patterns other than the uniform 0-flux (or $\pi$-flux) are stabilized in every plaquette. When the field is along [110] direction, 0 and $\pi$-fluxes coexist where the total fluxes penetrating the geometrical object enclosed by plaquettes should be quantized in unit of $2\pi$. However, for field along [111] direction, even more complicated flux patterns are manipulated to break the time-reversal symmetry. In this case, the fluxes through the minimal plaquette are adjusted as a function of the external field to minimize the ground state energy. The emergent Lorentz force then bends the spinon motion, which is reflected in the spinon band structure\cite{hirschberger2015large, gao2019topological, zhang2019topological}. As a consequence, the large thermal hall effect occurs from the topological spinon bands, $\kappa_{xy}/T\sim 4.6\times 10^{-3}\text{W}/(\text{K}^2\cdot\text{m})$ at low temperature below {1 \text{K}}. We discuss the relevant experiments and generalizations to the other rare-earth pyrochlore materials.

\label{sec:Lattice gauge theory with the Zeeman coupling}
{\textbf{\textit {Lattice gauge theory with the Zeeman coupling ---}}}
We first consider the nearest-neighbor pseudospin-1/2 model on a pyrochlore lattice capturing the essential features of the U(1) QSLs \cite{hermele2004pyrochlore}.
\bea
H_{\text{pseu}}&=&H_{\text{pseu}}^z+H_{\text{pseu}}^{\pm}
\nonumber \\
 &=&\sum_{\langle i,j\rangle}\Big\lbrace J_zS_i^zS_j^z-\frac{J_{\pm}}{2}(S_i^{+}S_j^{-}+S_i^{-}S_j^{+})\Big\rbrace, 
\label{eq:Ham}
\eea
where $J_z>0$ and $J_z\gg\vert J_{\pm}\vert$. Here, the pseudospin at each site $S_i^z$, $S_i^\pm$ is defined about their local $\hat{z}$ axis towards the center of a tetrahedron. The first term, Ising interaction determines the ground state manifold whose elements are called classical spin ice. The classical spin ice infers the spin configurations constrained by the ice rule, ``two in two out states".
Then, the second term serves as a perturbation lifting the degeneracy. The massive amounts of the degeneracy
is blended through the quantum tunneling, the second term in Eq. (\ref{eq:Ham}).

\begin{figure}[t]
  \begin{center}
  \includegraphics[width=0.6\linewidth]{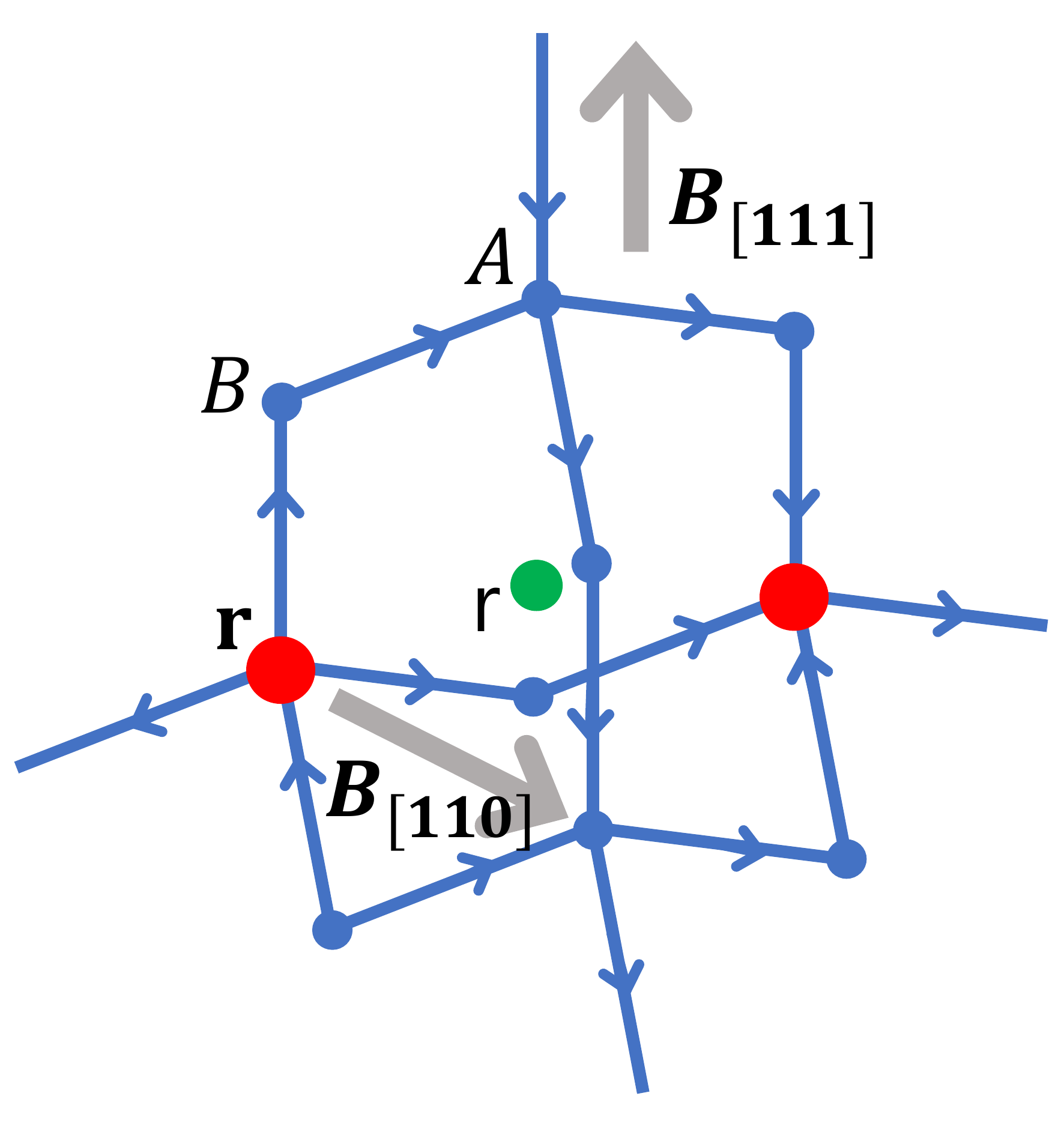}
     \caption{The diamond lattice with $A, B$ sublattices on which the spinon resides. On the link, the spin degrees of freedom live whose out-plane component $S_{\textbf{r}\textbf{r}}^z$ is marked as a blue arrow. The divergence $\sum_{\textbf{r}'}S_{\textbf{r}\textbf{r}}^z$ is identified with the gauge charge density (red circle) at $\textbf{r}$. The ring exchange is carried out on each 4 hexagonal plaquette enwrapping the dual diamond site $\mathsf{r}$ (green circle) where the magnetic monopole emitting the gauge flux resides. The external B-field is applied along [110]- or [111]-direction.}
    \label{fig:diamond}
  \end{center}
\end{figure}

The second term in Eq. \eqref{eq:Ham} creates a pair of bosonic spinon excitations on the center of the tetrahedra, the diamond lattice\cite{motrunich2002exotic, motrunich2005origin, castelnovo2008magnetic, jaubert2011magnetic}. In Fig. \ref{fig:diamond}, the diamond site is labelled by $\textbf{r}$ where the pyrochlore site $i$ is located at the center of the diamond bond connecting $\textbf{r}$ and $\textbf{r}'$. On the link, the spin variable $S_{\textbf{r}\textbf{r}'}^z$ can be thought as an electric field $E_{\textbf{r}\textbf{r}'}=\epsilon_{\textbf{r}\textbf{r}'}S_{\textbf{r}\textbf{r}'}^z$ where $\epsilon_{\textbf{r}\textbf{r}'}=-\epsilon_{\textbf{r}'\textbf{r}}=1$ when $\textbf{r}\in A$, $\textbf{r}'\in B$ and $\epsilon_{\textbf{r}\textbf{r}'}=0$ otherwise \cite{hermele2004pyrochlore}. Then the spinon excitation is mapped into the gauge charge $n_{\textbf{r}}=\sum_{\textbf{r}'\in\langle\textbf{r}\textbf{r}'\rangle}E_{\textbf{r}\textbf{r}'}$ obeying the Gauss law, and the Ising interaction in Eq. (\ref{eq:Ham}) becomes the charge repulsion $\frac{J_z}{2}\sum_{\textbf{r}}(n_{\textbf{r}})^2$.
 Similarly, the in-plane pseudospin $S_{\textbf{r}\textbf{r}'}^{+}=\Phi_{\textbf{r}}^{\dagger}e^{iA_{\textbf{r}\textbf{r}'}}\Phi_{\textbf{r}'}$ induces the spinon $\Phi_{\textbf{r}}$ to hop from $\textbf{r}'$ to $\textbf{r}$ with an emergent U(1) gauge field $A_{\textbf{r}\textbf{r}'}$ (mod $2\pi$). \cite{hermele2004pyrochlore}
  These gauge fields satisfy $[A_{\textbf{r}\textbf{r}'}$, $E_{\textbf{r}\textbf{r}'}]=i$ on each link.  
The exchange interaction in Eq. (\ref{eq:Ham}) represents the spinon propagation under the gauge field  
\bea
H_1=-\sum_{\langle\langle\textbf{r}\textbf{r}' \rangle\rangle}\Big(t_{\textbf{r}\textbf{r}'}\Phi_{\textbf{r}}^{\dagger}e^{iA_{\textbf{r}\textbf{r}'}}\Phi_{\textbf{r}} + \text{h.c.}\Big)-\mu\sum_{\textbf{r}}\Phi_{\textbf{r}}^{\dagger}\Phi_{\textbf{r}}
\label{eq:spinnon hopping}
\eea
where $t_{\textbf{r}\textbf{r}'}=\frac{J_{\pm}}{2}$ and $A_{\textbf{r}\textbf{r}'}=A_{\textbf{r}\textbf{r}''}+A_{\textbf{r}''\textbf{r}'}$. For convenience, the spinon charge is set to be $q=1$ in unit of $\hbar=c=1$.
The spinon spectrum is manifest in the band structure Eq. (\ref{eq:spinnon hopping}) above a finite energy gap $\vert \mu\vert \sim J_z$. 
Generally, the long-range spinon hopping other than Eq. (\ref{eq:spinnon hopping}) can be built in the products $S_{\textbf{r}\textbf{a}}^{+}S_{\textbf{a}\textbf{b}}^{-}\ldots S_{\textbf{a}'\textbf{r}'}^{-}$. 

By integrating out the gapped spinons in Eq. (\ref{eq:spinnon hopping}), the $3^{\text{rd}}$ order perturbation gives the compact U(1) gauge theory with the ring exchange around the hexagonal plaquette\cite{hermele2004pyrochlore}.
\bea
H_{\text{eff}}^{(3)}=\frac{U}{2}\sum_{\text{link}}E_{\textbf{r}\textbf{r}'}^2-\sum_{\text{plaq}}g_p\text{cos}(\triangledown\times A_{\textbf{r}\textbf{r}'}),
\label{eq:3rd pert}
\eea
where $g_p=3J_{\pm}^3/2J_{z}^2$ is the coupling constant for the ring exchange. The first term is included to enforce the discreteness of the pseudospin for large $U>0$ and the second term denotes the lattice curl around the hexagonal plaquette. 
In addition to the spinons we noted, two more emergent excitations exist, magnetic monopole and photon at lower energy\cite{hermele2004pyrochlore}.

In U(1) QSL, the gapped spinons are deconfined and propagate in the dual lattice. 
Comparing Eq. (\ref{eq:spinnon hopping}) and (\ref{eq:3rd pert}), it is obvious that the gauge flux stabilized by the coupling constant $g_p$ is decisive for the spinon band structure\cite{wen2002quantum, lee2012generic, essin2013classifying, chen2017spectral}. With unfrustrated $J_{\pm}>0$, all plaquettes prefer the 0-flux and the spinon band structure is the same as the one of the diamond lattice without any gauge field. In the frustrated case $J_{\pm}<0$, the unit cell is doubly enlarged to stabilize the $\pi$-flux with the line degeneracy in the band structure\cite{lee2012generic}.

Now we take into account the Zeeman term by applying the external magnetic field $\textbf{B}$. 
\bea
H_{\text{Zeeman}}&=&H_{\text{Z}}^z+H_{\text{Z}}^{\pm}
\nonumber\\
&=&-\sum_{i}h_i^zS_i^z-\sum_{i}(h_i^xS_i^x+h_i^yS_i^y),
\label{eq:Zeeman}
\eea
where the Bohr magneton $\mu_B$ and the $g$-factor are absorbed into the definition $h=\mu_B g\vert\textbf{B}\vert$. The subscript $i$ is inserted to remind that the relative angle between the local axis and the B-field depending on four sublattices in a pyrochlore lattice.
We focus on the small field regime where the U(1) QSL is still stable and the role of magnetic fields in the ring exchange terms modifying U(1) gauge fields.
The second term $H_{\text{Z}}^{\pm}$ in Eq. \eqref{eq:Zeeman} influences the effective models in Eqs. (\ref{eq:spinnon hopping}) and (\ref{eq:3rd pert}). 
We also note that the first term in Eq. \eqref{eq:Zeeman} does not contribute to the lowest order correction upto fourth order in fields, thus is ignored.

The spin-flip term in Eq. (\ref{eq:Zeeman}) calls for the nearest neighbor spinon hopping in Eq. (\ref{eq:spinnon hopping}). 
And it modifies the coupling constant $g_p$ through the ring exchange, resulting in staggered flux patterns other than uniform 0-flux or $\pi$-flux. As shown in Fig. \ref{fig:diamond}, there are 4 hexagonal plaquette completely wrapping the dual diamond lattice where the source of the gauge flux resides. Within the moderate strength of $h$, some of the plaquettes prefer 0-flux while the others prefer $\pi$-flux among 4 different faces. When two among four plaquettes favor 0-flux and others favor $\pi$-flux, the spinon band remains topologically trivial satisfying the net flux penetrating the 4 plaquettes quantized in unit of $2\pi$.
However, there are cases where one or three plaquettes prefer $\pi$-flux among four plaquettes. In these cases, the net flux is not quantized in unit of $2\pi$ and the U(1) QSL description breaks down. The validity can be restored by shaving off the flux from the preferred one. It is implemented to minimize the magnetic energy similar to the uniform flux cases. Such flux is adjusted in a continuous fashion by the external field and the topological spinon bands arise with broken time reversal symmetry. 

\begin{figure}[t]
\subfloat[]{\label{fig:110}\includegraphics[width=0.5\textwidth]{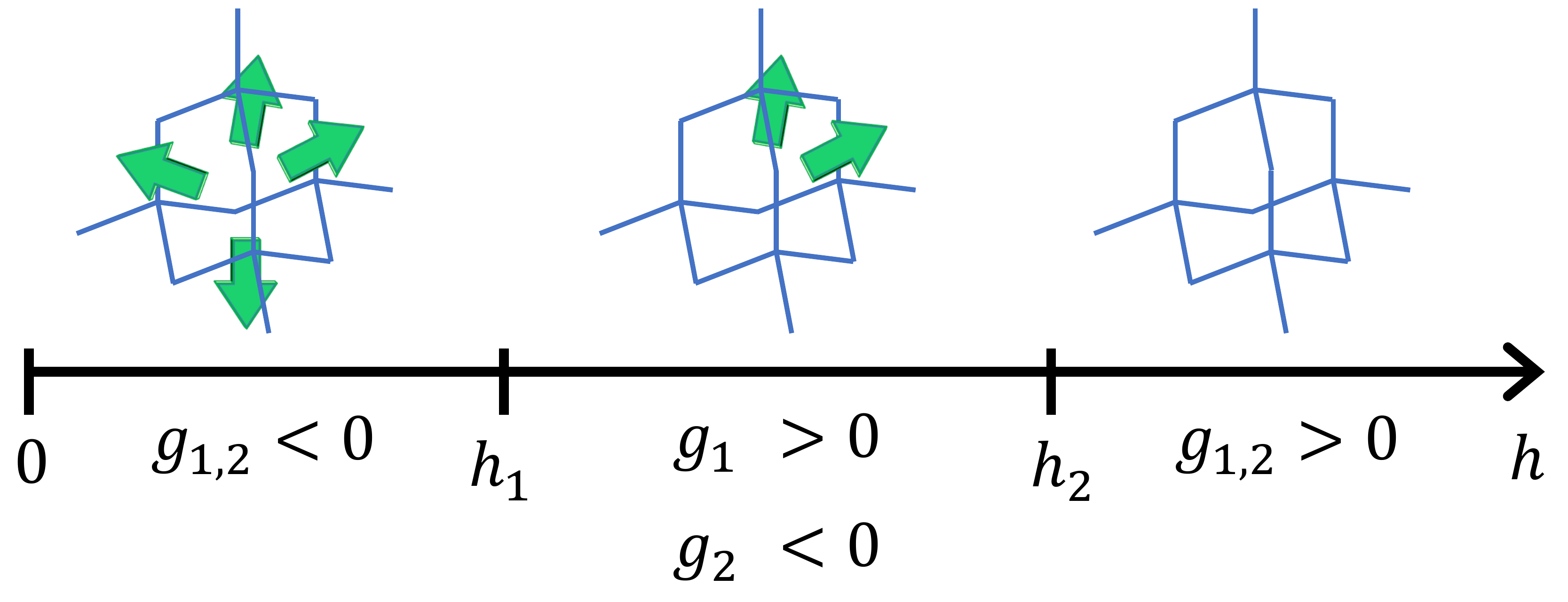}}
\nonumber\\
\subfloat[]{\label{fig:111}\includegraphics[width=0.5\textwidth]{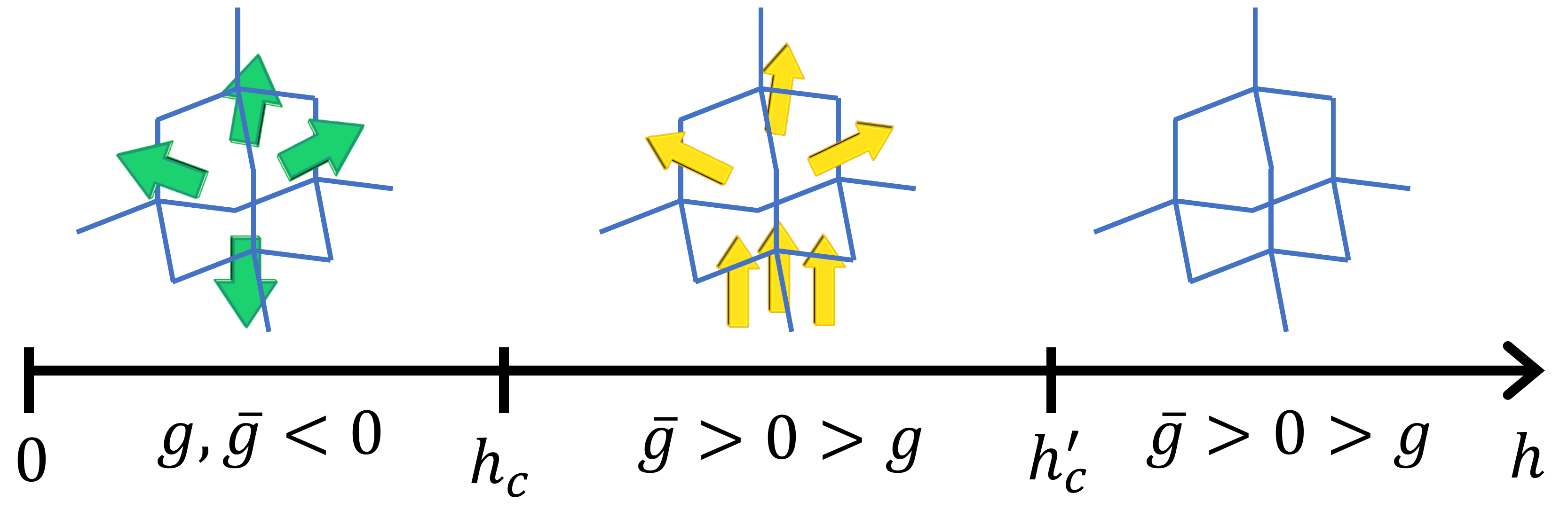}}
\caption{Schematic phase diagrams with the signs of the coupling constant $g_p$ when the exchange is frustrated $J_{\pm}<0$ and the B-field is applied along (a) [110]- and (b) [111]-directions. The blank, green and yellow arrows are identified with 0-, $\pi$- and continuous $\mathcal{B}$-fluxes respectively. The phase diagrams are continued until $h$ polarizes the system out of U(1) QSLs.}
\label{fig:phasediagram}
\end{figure}

\label{sec: Perturbation g}
{\textbf{\textit {Perturbation of the coupling $g$ in fields ---}}} We concrete our argument in the presence of the B-field along [110]- and [111]-directions. Since the ring exchange consists of the 6 subsequent spin flips, the lowest order where $h$ comes in is 4. 
After the $4^{\text{th}}$-order perturbation (Supplementary Information), the effective Hamiltonian Eq. (\ref{eq:3rd pert}) is obtained with the modified coupling constant. When the B-field is along [110]-direction,
\bea
g_1^{[110]}=\frac{3J_{\pm}^3}{2J_z^2}+\frac{5J_{\pm}^2h^2}{4J_z^3}
\nonumber\\
g_2^{[110]}=\frac{3J_{\pm}^3}{2J_z^2}+\frac{J_{\pm}^2h^2}{J_z^3}
\label{eq:110}
\eea
where $g_1^{[110]}\;(g_2^{[110]})$ is the coupling constant on the plaquette parallel (oblique) to the B-field direction. The difference in the corrections is due to the relative angle between the plaquette and the external B-field.
When the exchange is unfrustrated ($J^{\pm}>0$), the B-field just enhances the stability of the 0-flux U(1) QSLs. Although higher order perturbation $\sim h^4J_{\pm}^2/J_z^5$ may introduce the terms with the opposite signs, it will only renormalize Eq. (\ref{eq:110}) unless the B-field is relatively large $h \sim (J_z^3J_{\pm})^{1/4}$.
However, in the frustrated case ($J^{\pm}<0$), they reverse the coupling constant signs as the B-field approches $h\sim \vert J_zJ_{\pm}\vert^{1/2}$. (Fig. \ref{fig:110}) At small field $h<h_1=1.10\vert J_zJ_{\pm}\vert^{1/2}$, the coupling constants $g_{1(2)}^{[110]}$ are negative implying the $\pi$-flux phase. With further increasing $h$, all constants in Eq. (\ref{eq:110}) switch the signs beyond $h>h_2=1.22\vert J_zJ_{\pm}\vert^{1/2}$ and the 0-flux U(1) QSL is stabilized before polarization for large $h$. Inbetween these two transition points $h_1<h<h_2$, only two plaquettes among 4 faces keep the negative constants $g_{1}^{[110]}>0, g_{2}^{[110]}<0$. In this regime, the ground state no longer stabilizes the uniform gauge fluxes. Rather, two 0-fluxes and two $\pi$-fluxes among 4 faces are stabilized whose net flux is well quantized. Although the gapless photon is insensitive to small $h$, the spinon band structure changes passing the transition points $h_{1,2}$. All band structures in three regimes are topologically trivial since the 0- and $\pi$-fluxes respect the time-reversal symmetry. 

Unlike the [110] field case, the field along [111] direction results in frustrated U(1) gauge fluxes and non-trivial spinon band structure is stabilized. 
In this case, the corrections to the coupling constants are,
\bea
g&=&\frac{3J_{\pm}^3}{2J_z^2}
\nonumber\\
\bar{g}&=&\frac{3J_{\pm}^3}{2J_z^2}+\frac{10}{9}\frac{J_{\pm}^2h^2}{J_z^3},
\label{eq:111}
\eea
where the superscript [111] is omitted for convenience. 
Here, $g$ correspond to the plaquette perpendicular to the B-field and $\bar{g}$ to the other 3 tilted faces. Similar to the [110] case, the unfrustrated exchange $J_\pm$ is not affected by the applied field. For $J_{\pm}<0$, both constants are negative below $h_c=1.16\vert J_zJ_{\pm}\vert^{1/2}$ preferring the uniform $\pi$-flux. When the B-field reaches $h=h_c$, then $\bar{g}=0$ and the ring exchanges are completely suppressed  without the perpendicular kagome plane. At this transition point, the U(1) QSL description fails due to the instanton effect in space-time 2+1 dimension. \cite{polyakov1975interaction}

When $h>h_c$, the phase with staggered flux contains the plaquette 
preferring the $\pi$-flux perpendicular to [111]-direction and 0-flux on the others. The obstacle is that the net flux penetrating the 4 different faces is not quantized in unit of $2\pi$. The flux quantization is legitimated by shaving off the preferred flux in each plaquette.
Thus we minimize the total magnetic energy instead of each plaquette.
\bea
H_{\text{eff}}^{[111]} &\sim &  \vert g\vert\text{cos}(2\pi -3\mathcal{B})-3\bar{g}\text{cos}(\mathcal{B})
\nonumber\\
&=&4\vert g\vert \text{cos}^3(\mathcal{B})-3(\vert g\vert + \bar{g})\text{cos}(\mathcal{B})
\label{eq:111staggered}
\eea
where $\mathcal{B}$ is the sheared flux through the 3 tilted plaquettes.
Since Eq. (\ref{eq:111staggered}) is a polynomial of $\text{cos}({\mathcal{B}})$ and the leading coefficient is positive, the sheared flux is in the vicinity of 0 rather than $\pi$ as it should be. Minimizing Eq. (\ref{eq:111staggered}), the optimized flux is
\bea
\frac{dH_{\text{eff}}^{[111]}}{d\mathcal{B}}\Big\vert_{\mathcal{B}=\mathcal{\bar{B}}}=0,\quad \text{cos}(\bar{\mathcal{B}})=\sqrt{\frac{\vert g\vert + \bar{g}}{4\vert g\vert}}
\label{eq:Optimized B}
\eea
This allows the gauge flux $\mathcal{B}$ other than standard 0- and $\pi$-fluxes. When $h$ is larger than $h'_{c}=2.32(J_zJ_{\pm})^{1/2}$, then $\bar{g}>3|g| $ and Eq. (\ref{eq:Optimized B}) has no solution for $\bar{\mathcal{B}}$. In this regime, all plaquettes are enforced to trap the 0-flux evenly the negative $g<0$ one.
Between the transition points $h_c<h<h'_c$, the solution $\bar{\mathcal{B}}$ of Eq. (\ref{eq:Optimized B}) is stabilized through the 3 tilted plaquettes with  $3\bar{\mathcal{B}}$-flux perpendicular plaquette. Here, $\bar{\mathcal{B}}$ is a continuous function of the strength $h$ with the range $0\leq\bar{\mathcal{B}}\leq\pi/3$ since $\bar{g}\geq 0$.


\label{sec: staggered}
{\textbf{\textit {Staggered flux phase and thermal Hall effect ---}}}
In the intermediate field $h_c<h<h'_c$ along [111] direction, the time-reversal symmetry is broken for generic staggered flux $\bar{\mathcal{B}}$.
Since the spinon experiences the emergent gauge flux $\bar{\mathcal{B}}$ (and $3\bar{\mathcal{B}}$), the emergent Lorentz force affects the spinon kinetics \cite{gao2019topological, zhang2019topological}. This is manifest in the spinon band structure which may exhibit finite Chern numbers. Even the arbitrary strength of $h$ may lead to the incommensurate $\mathcal{B}$ with the Hofstadter butterfly, the non-trivial band structure is typically discernible in this regime\cite{hofstadter1976energy, bernevig2013topological}. Above the temperature about the spinon gap, this leads to the thermal hall effect stimulated by the Berry curvature $\Omega_{k}$\cite{katsura2010theory, hirschberger2015large, murakami2016thermal}. In this phase, the spinon subject to the magnetic field carries a heat current perpendicular to the temperature gradient.
The proportional coefficient $\kappa_{xy}$ is sensitive to the Berry curvature in which the spinons are thermally populated. 
The parameters $J_{\pm}$ and $h$ implicitly adjust the band structure resulting in the overall weight in $\kappa_{xy}$.
\begin{figure}[b]
  \begin{center}{}
  \includegraphics[width=0.45\textwidth]{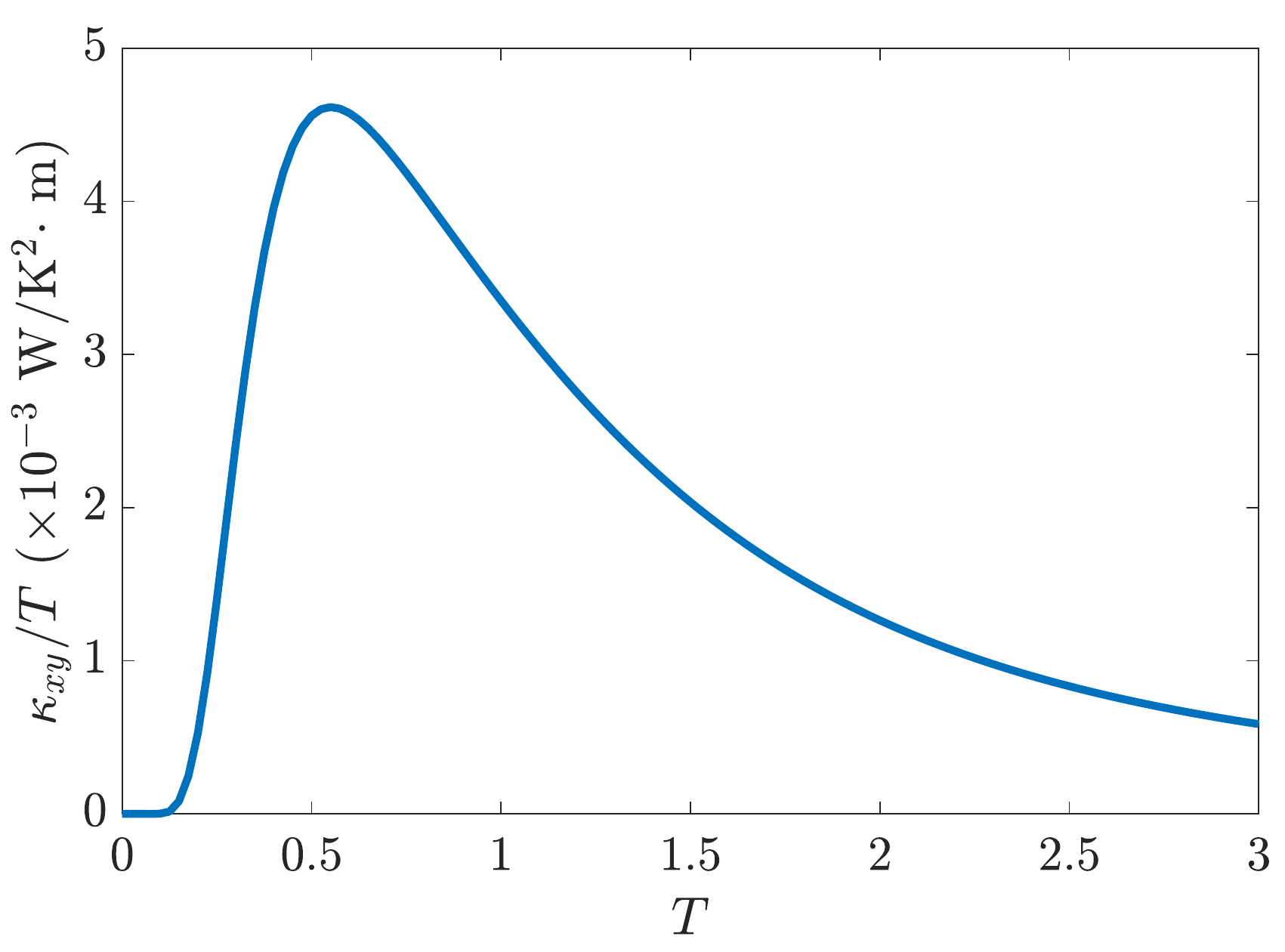}
\caption{
Numerical plot of the thermal hall coefficient $\kappa_{xy}/T$ as the temperature $T$ changes when $\vert g\vert =\bar{g}$ and $\bar{\mathcal{B}}=\pi/4$. 
For simplicity, it is estimated with $t_{\textbf{r}\textbf{r}'}=1$ for nearest neighbors only and the chemical potential $\mu=-4$ in Eq. (\ref{eq:spinnon hopping}).
In the vertical axis, the unit is evaluated with respect to the diamond lattice constant $\sim 4.3\text{\AA}$\cite{jaubert2011magnetic}. In the horizontal axis, the temperature is in unit of $t_{\textbf{r}\textbf{r}'}\sim J_{\pm}$.}
\label{fig:Numericalplot}
\end{center}
\end{figure}

We numerically evaluate the spinon contribution to the thermal hall conductivity $\kappa_{xy}(T)$. Nonvanishing coefficient $\kappa_{xy}$ will signal the relevant evidence for detecting the staggered flux phase. For concreteness, we consider a simple example reasonable for  $|J_{\pm}/J_z|<1$, $t_{\textbf{r}\textbf{r}'}=1$ and $\mu=-4$ keeping only the nearest neighbor hopping due to applying magnetic field in Eq. (\ref{eq:spinnon hopping}).
We set the gauge field $\vert g \vert = \bar{g}$ and $\mathcal{B}=\pi/4$ in Eq. (\ref{eq:Optimized B}). With an appropriate gauge fixing (See Supplementary Information for details), the unit cell of the diamond lattice is enlarged 16 times and the Brillouin zone is folded to yield 32 spinon bands. Based on the band structure, the thermal hall coefficient $\kappa_{xy}(T)/T$ is evaluated where \cite{luttinger1964theory, smrcka1977transport, matsumoto2011theoretical, matsumoto2014thermal, murakami2016thermal},
\bea
\kappa_{xy}(T)\!=\!\frac{k_B^2T}{\hbar}\sum_{n=1}^{32}\int\frac{d^3k}{(2\pi)^3}\Big\lbrace c_2[g(\epsilon_{n,\textbf{k}})]\!-\!\frac{\pi^2}{3} \Big\rbrace\Omega_{n,\textbf{k}}.
\label{eq:ThermalHall}
\eea
Here, $\Omega_{n,\textbf{k}}$ is the Berry curvature at $\textbf{k}$ on the $n\text{-th}$ band. The weight  
$c_2(x)=(1+x)[\text{ln}((1+x)/x)]^2-(\text{ln}x)^2-2\text{Li}_2(-x)$ with a polylogarithmic function $\text{Li}_2(x)$ inherit the role of the Bose distribution $g(\epsilon_{n,\textbf{k}})=1/(e^{\epsilon_{n,\textbf{k}}/k_BT}-1)$. 
Fig. \ref{fig:Numericalplot} shows the thermal hall conductivity $\kappa_{xy}/T$ as a function of temperature $T$. 
The partially occupied spinon bands give rise to the dominant behavior of $\kappa_{xy}$. As the temperature gradually increases, it grows from $\kappa_{xy}/T=0$ towards the peak due to the thermal population of the non-trivial spinon bands. It is noteworthy that the maximum thermal hall signal $\kappa_{xy}/T\sim 4.6\times 10^{-3}\text{W}/(\text{K}^2\cdot\text{m})$ near  $T\sim 0.5 J_\pm$ is huge and thus is expected to be accessible in experiments\cite{hirschberger2015large, hirschberger2015thermal, kasahara2018majorana, yu2018ultralow, hirschberger2019enhanced}. 
There also exist the contributions from the magnetic monopole and gapless photon\cite{zhang2019topological}. However, these contributions exist at very low temperature and are thought to be the substracted background in overall weight near $T=0\text{K}$. When $T$ approaches the spinon gap $\sim J_z$, those bands are almost fully occupied and the background fades out. 
For further increasing temperature, it starts to decrease due to the population in the higher bands with opposite signs of the Chern numbers. It vanishes as $T$ goes further when all spinon bands are trivially occupied.





\label{sec: Conclusion}
{\textbf{\textit {Conclusion ---}}} 
We briefly comment the effect of additional exchange interactions.
Despite our analysis is based on the leading terms Eqs.  
(\ref{eq:Ham}) and (\ref{eq:Zeeman}), it can be generalized including additional types of exchange interactions. Generically, the symmetry allows other exchanges such as 
$\sim J_{\pm \pm}S^{\pm}_iS^{\pm}_{j}/2$ and $\sim J_{\pm z}S^{\pm}_iS^z_{j}/2$\cite{ross2011quantum, savary2012coulombic, lee2012generic, hao2014bosonic}.
{If the former term $\sim J_{\pm \pm}S^{\pm}_1S^{\pm}_{2}/2$ is embodied in the ring exchange, the spins at site $1$ or $2$ is required to be flipped at least three times subsequently, thus it does not contribute as a dominant term but plays a role beyond the 4th order perturbation.}
Whereas, the latter term $\sim J_{\pm z}S^{\pm}_iS^z_{j}/2$ assembles the Zeeman coupling Eq. (\ref{eq:Zeeman}) and modifies the coupling constants Eqs. (\ref{eq:110}) and (\ref{eq:111}) in the same manner. The lowest order correction results in [Appendix B]
$\frac{3J_{\pm}^3}{2J_z^2} \rightarrow \frac{3J_{\pm}^3}{2J_z^2}+9\frac{J_{\pm}^2J_{\pm z}^2}{J_z^3}
$
in Eqs. (\ref{eq:110}) and (\ref{eq:111}). Likewise the external magnetic field, the frustrated plaquettes $g_p<0$ reverse its sign at $J_{\pm z}=0.41\vert J_{\pm}J_z \vert^{1/2}$ in the absence of the B-field. Thus, below the critical value $J_{\pm z}<0.41\vert J_{\pm}J_z \vert^{1/2}$, the interaction  $\sim J_{\pm z}S^{\pm}_iS^z_{j}/2$ induces smaller critical fields in Fig. (\ref{fig:phasediagram}). This enhances the stability of QSLs even turning on B-field and one expects the observable $\kappa_{xy}$ in the regime $h_c<h<h_c'=2h_c$ where $h_c \rightarrow 1.16\vert J_{z}J_{\pm}-6J_{\pm z}^2\vert^{1/2}$ decreases.

In this paper, we study the pyrochlore U(1) QSL applied to the magnetic field. The Zeeman coupling modifies the coupling constant in the emergent U(1) gauge theory, especially critically affects to the spinon spectrum. The spinon kinetics experiencing the gauge field is controlled by the strength of the B-field. With the B-field along [110]-direction, the gauge flux choose 0 or $\pi$-flux which preserves the time-reversal symmetry. Meanwhile, with the B-field along [111]-direction, the preferred fluxes covering the magnetic source are frustrated in the intermediate regime. This enforces the flux to be shaved off continuously, or evenly trap the unfavored flux. In this regime, the thermal hall measurement is an accessible setup to convince the change in emergent gauge structure at low energy. 
It turns out that spinon bands with frustrated fluxes become topologically non-trivial and results in large thermal hall effect $\kappa_{xy}/T\sim 4.6\times 10^{-3}\text{W}/(\text{K}^2\cdot\text{m})$ at low temperature below 1K.
 Our argument can be generally applicable to pyrochlore U(1) QSLs and associated thermal Hall measurement in fields would advocate the U(1) gauge structure and existence of emergent fractional quasiparticles in QSLs. 
Based on our study, the strain effect in rare-earth pyrochlores with non-Kramer doublet can be another interesting point as relevant future work. 

\label{sec: Acknowledgments}
{\textbf{\textit {Acknowledgments ---}}}
We thank Leon Balents, Jeffrey G. Rau, Jung Hoon Han, Subhro Bhattacharjee and Nic Shannon  for many valuable discussions. This work is supported by the KAIST startup, National Research Foundation Grant (NRF-2017R1A2B4008097) and in part by the National Science Foundation under Grant No. NSF PHY-1748958.

\bibliography{ref}

\begin{thebibliography}{64}%
\makeatletter
\providecommand \@ifxundefined [1]{%
 \@ifx{#1\undefined}
}%
\providecommand \@ifnum [1]{%
 \ifnum #1\expandafter \@firstoftwo
 \else \expandafter \@secondoftwo
 \fi
}%
\providecommand \@ifx [1]{%
 \ifx #1\expandafter \@firstoftwo
 \else \expandafter \@secondoftwo
 \fi
}%
\providecommand \natexlab [1]{#1}%
\providecommand \enquote  [1]{``#1''}%
\providecommand \bibnamefont  [1]{#1}%
\providecommand \bibfnamefont [1]{#1}%
\providecommand \citenamefont [1]{#1}%
\providecommand \href@noop [0]{\@secondoftwo}%
\providecommand \href [0]{\begingroup \@sanitize@url \@href}%
\providecommand \@href[1]{\@@startlink{#1}\@@href}%
\providecommand \@@href[1]{\endgroup#1\@@endlink}%
\providecommand \@sanitize@url [0]{\catcode `\\12\catcode `\$12\catcode
  `\&12\catcode `\#12\catcode `\^12\catcode `\_12\catcode `\%12\relax}%
\providecommand \@@startlink[1]{}%
\providecommand \@@endlink[0]{}%
\providecommand \url  [0]{\begingroup\@sanitize@url \@url }%
\providecommand \@url [1]{\endgroup\@href {#1}{\urlprefix }}%
\providecommand \urlprefix  [0]{URL }%
\providecommand \Eprint [0]{\href }%
\providecommand \doibase [0]{http://dx.doi.org/}%
\providecommand \selectlanguage [0]{\@gobble}%
\providecommand \bibinfo  [0]{\@secondoftwo}%
\providecommand \bibfield  [0]{\@secondoftwo}%
\providecommand \translation [1]{[#1]}%
\providecommand \BibitemOpen [0]{}%
\providecommand \bibitemStop [0]{}%
\providecommand \bibitemNoStop [0]{.\EOS\space}%
\providecommand \EOS [0]{\spacefactor3000\relax}%
\providecommand \BibitemShut  [1]{\csname bibitem#1\endcsname}%
\let\auto@bib@innerbib\@empty
\bibitem [{\citenamefont {Anderson}(1973)}]{anderson1973resonating}%
  \BibitemOpen
  \bibfield  {author} {\bibinfo {author} {\bibfnamefont {P.~W.}\ \bibnamefont
  {Anderson}},\ }\href@noop {} {\bibfield  {journal} {\bibinfo  {journal}
  {Materials Research Bulletin}\ }\textbf {\bibinfo {volume} {8}},\ \bibinfo
  {pages} {153} (\bibinfo {year} {1973})}\BibitemShut {NoStop}%
\bibitem [{\citenamefont {Anderson}(1987)}]{anderson1987resonating}%
  \BibitemOpen
  \bibfield  {author} {\bibinfo {author} {\bibfnamefont {P.~W.}\ \bibnamefont
  {Anderson}},\ }\href@noop {} {\bibfield  {journal} {\bibinfo  {journal}
  {science}\ }\textbf {\bibinfo {volume} {235}},\ \bibinfo {pages} {1196}
  (\bibinfo {year} {1987})}\BibitemShut {NoStop}%
\bibitem [{\citenamefont {Fradkin}(2013)}]{fradkin2013field}%
  \BibitemOpen
  \bibfield  {author} {\bibinfo {author} {\bibfnamefont {E.}~\bibnamefont
  {Fradkin}},\ }\href@noop {} {\emph {\bibinfo {title} {Field theories of
  condensed matter physics}}}\ (\bibinfo  {publisher} {Cambridge University
  Press},\ \bibinfo {year} {2013})\BibitemShut {NoStop}%
\bibitem [{\citenamefont {Balents}(2010)}]{balents2010spin}%
  \BibitemOpen
  \bibfield  {author} {\bibinfo {author} {\bibfnamefont {L.}~\bibnamefont
  {Balents}},\ }\href@noop {} {\bibfield  {journal} {\bibinfo  {journal}
  {Nature}\ }\textbf {\bibinfo {volume} {464}},\ \bibinfo {pages} {199}
  (\bibinfo {year} {2010})}\BibitemShut {NoStop}%
\bibitem [{\citenamefont {Savary}\ and\ \citenamefont
  {Balents}(2016)}]{savary2016quantum}%
  \BibitemOpen
  \bibfield  {author} {\bibinfo {author} {\bibfnamefont {L.}~\bibnamefont
  {Savary}}\ and\ \bibinfo {author} {\bibfnamefont {L.}~\bibnamefont
  {Balents}},\ }\href@noop {} {\bibfield  {journal} {\bibinfo  {journal}
  {Reports on Progress in Physics}\ }\textbf {\bibinfo {volume} {80}},\
  \bibinfo {pages} {016502} (\bibinfo {year} {2016})}\BibitemShut {NoStop}%
\bibitem [{\citenamefont {Zhou}\ \emph {et~al.}(2017)\citenamefont {Zhou},
  \citenamefont {Kanoda},\ and\ \citenamefont {Ng}}]{zhou2017quantum}%
  \BibitemOpen
  \bibfield  {author} {\bibinfo {author} {\bibfnamefont {Y.}~\bibnamefont
  {Zhou}}, \bibinfo {author} {\bibfnamefont {K.}~\bibnamefont {Kanoda}}, \ and\
  \bibinfo {author} {\bibfnamefont {T.-K.}\ \bibnamefont {Ng}},\ }\href@noop {}
  {\bibfield  {journal} {\bibinfo  {journal} {Reviews of Modern Physics}\
  }\textbf {\bibinfo {volume} {89}},\ \bibinfo {pages} {025003} (\bibinfo
  {year} {2017})}\BibitemShut {NoStop}%
\bibitem [{\citenamefont {Lacroix}\ \emph {et~al.}(2011)\citenamefont
  {Lacroix}, \citenamefont {Mendels},\ and\ \citenamefont
  {Mila}}]{lacroix2011introduction}%
  \BibitemOpen
  \bibfield  {author} {\bibinfo {author} {\bibfnamefont {C.}~\bibnamefont
  {Lacroix}}, \bibinfo {author} {\bibfnamefont {P.}~\bibnamefont {Mendels}}, \
  and\ \bibinfo {author} {\bibfnamefont {F.}~\bibnamefont {Mila}},\ }\href@noop
  {} {\emph {\bibinfo {title} {Introduction to frustrated magnetism: materials,
  experiments, theory}}},\ Vol.\ \bibinfo {volume} {164}\ (\bibinfo
  {publisher} {Springer Science \& Business Media},\ \bibinfo {year}
  {2011})\BibitemShut {NoStop}%
\bibitem [{\citenamefont {Wynn}\ \emph {et~al.}(2001)\citenamefont {Wynn},
  \citenamefont {Bonn}, \citenamefont {Gardner}, \citenamefont {Lin},
  \citenamefont {Liang}, \citenamefont {Hardy}, \citenamefont {Kirtley},\ and\
  \citenamefont {Moler}}]{wynn2001limits}%
  \BibitemOpen
  \bibfield  {author} {\bibinfo {author} {\bibfnamefont {J.}~\bibnamefont
  {Wynn}}, \bibinfo {author} {\bibfnamefont {D.}~\bibnamefont {Bonn}}, \bibinfo
  {author} {\bibfnamefont {B.}~\bibnamefont {Gardner}}, \bibinfo {author}
  {\bibfnamefont {Y.-J.}\ \bibnamefont {Lin}}, \bibinfo {author} {\bibfnamefont
  {R.}~\bibnamefont {Liang}}, \bibinfo {author} {\bibfnamefont
  {W.}~\bibnamefont {Hardy}}, \bibinfo {author} {\bibfnamefont
  {J.}~\bibnamefont {Kirtley}}, \ and\ \bibinfo {author} {\bibfnamefont
  {K.}~\bibnamefont {Moler}},\ }\href@noop {} {\bibfield  {journal} {\bibinfo
  {journal} {Physical review letters}\ }\textbf {\bibinfo {volume} {87}},\
  \bibinfo {pages} {197002} (\bibinfo {year} {2001})}\BibitemShut {NoStop}%
\bibitem [{\citenamefont {Harris}\ \emph {et~al.}(1997)\citenamefont {Harris},
  \citenamefont {Bramwell}, \citenamefont {McMorrow}, \citenamefont {Zeiske},\
  and\ \citenamefont {Godfrey}}]{harris1997geometrical}%
  \BibitemOpen
  \bibfield  {author} {\bibinfo {author} {\bibfnamefont {M.~J.}\ \bibnamefont
  {Harris}}, \bibinfo {author} {\bibfnamefont {S.}~\bibnamefont {Bramwell}},
  \bibinfo {author} {\bibfnamefont {D.}~\bibnamefont {McMorrow}}, \bibinfo
  {author} {\bibfnamefont {T.}~\bibnamefont {Zeiske}}, \ and\ \bibinfo {author}
  {\bibfnamefont {K.}~\bibnamefont {Godfrey}},\ }\href@noop {} {\bibfield
  {journal} {\bibinfo  {journal} {Physical Review Letters}\ }\textbf {\bibinfo
  {volume} {79}},\ \bibinfo {pages} {2554} (\bibinfo {year}
  {1997})}\BibitemShut {NoStop}%
\bibitem [{\citenamefont {Canals}\ and\ \citenamefont
  {Lacroix}(1998)}]{canals1998pyrochlore}%
  \BibitemOpen
  \bibfield  {author} {\bibinfo {author} {\bibfnamefont {B.}~\bibnamefont
  {Canals}}\ and\ \bibinfo {author} {\bibfnamefont {C.}~\bibnamefont
  {Lacroix}},\ }\href@noop {} {\bibfield  {journal} {\bibinfo  {journal}
  {Physical Review Letters}\ }\textbf {\bibinfo {volume} {80}},\ \bibinfo
  {pages} {2933} (\bibinfo {year} {1998})}\BibitemShut {NoStop}%
\bibitem [{\citenamefont {Bramwell}\ \emph {et~al.}(2001)\citenamefont
  {Bramwell}, \citenamefont {Harris}, \citenamefont {Den~Hertog}, \citenamefont
  {Gingras}, \citenamefont {Gardner}, \citenamefont {McMorrow}, \citenamefont
  {Wildes}, \citenamefont {Cornelius}, \citenamefont {Champion}, \citenamefont
  {Melko} \emph {et~al.}}]{bramwell2001spin}%
  \BibitemOpen
  \bibfield  {author} {\bibinfo {author} {\bibfnamefont {S.}~\bibnamefont
  {Bramwell}}, \bibinfo {author} {\bibfnamefont {M.}~\bibnamefont {Harris}},
  \bibinfo {author} {\bibfnamefont {B.}~\bibnamefont {Den~Hertog}}, \bibinfo
  {author} {\bibfnamefont {M.}~\bibnamefont {Gingras}}, \bibinfo {author}
  {\bibfnamefont {J.}~\bibnamefont {Gardner}}, \bibinfo {author} {\bibfnamefont
  {D.}~\bibnamefont {McMorrow}}, \bibinfo {author} {\bibfnamefont
  {A.}~\bibnamefont {Wildes}}, \bibinfo {author} {\bibfnamefont
  {A.}~\bibnamefont {Cornelius}}, \bibinfo {author} {\bibfnamefont
  {J.}~\bibnamefont {Champion}}, \bibinfo {author} {\bibfnamefont
  {R.}~\bibnamefont {Melko}},  \emph {et~al.},\ }\href@noop {} {\bibfield
  {journal} {\bibinfo  {journal} {Physical Review Letters}\ }\textbf {\bibinfo
  {volume} {87}},\ \bibinfo {pages} {047205} (\bibinfo {year}
  {2001})}\BibitemShut {NoStop}%
\bibitem [{\citenamefont {Nakatsuji}\ \emph {et~al.}(2006)\citenamefont
  {Nakatsuji}, \citenamefont {Machida}, \citenamefont {Maeno}, \citenamefont
  {Tayama}, \citenamefont {Sakakibara}, \citenamefont {Van~Duijn},
  \citenamefont {Balicas}, \citenamefont {Millican}, \citenamefont {Macaluso},\
  and\ \citenamefont {Chan}}]{nakatsuji2006metallic}%
  \BibitemOpen
  \bibfield  {author} {\bibinfo {author} {\bibfnamefont {S.}~\bibnamefont
  {Nakatsuji}}, \bibinfo {author} {\bibfnamefont {Y.}~\bibnamefont {Machida}},
  \bibinfo {author} {\bibfnamefont {Y.}~\bibnamefont {Maeno}}, \bibinfo
  {author} {\bibfnamefont {T.}~\bibnamefont {Tayama}}, \bibinfo {author}
  {\bibfnamefont {T.}~\bibnamefont {Sakakibara}}, \bibinfo {author}
  {\bibfnamefont {J.}~\bibnamefont {Van~Duijn}}, \bibinfo {author}
  {\bibfnamefont {L.}~\bibnamefont {Balicas}}, \bibinfo {author} {\bibfnamefont
  {J.}~\bibnamefont {Millican}}, \bibinfo {author} {\bibfnamefont
  {R.}~\bibnamefont {Macaluso}}, \ and\ \bibinfo {author} {\bibfnamefont
  {J.~Y.}\ \bibnamefont {Chan}},\ }\href@noop {} {\bibfield  {journal}
  {\bibinfo  {journal} {Physical review letters}\ }\textbf {\bibinfo {volume}
  {96}},\ \bibinfo {pages} {087204} (\bibinfo {year} {2006})}\BibitemShut
  {NoStop}%
\bibitem [{\citenamefont {Molavian}\ \emph {et~al.}(2007)\citenamefont
  {Molavian}, \citenamefont {Gingras},\ and\ \citenamefont
  {Canals}}]{molavian2007dynamically}%
  \BibitemOpen
  \bibfield  {author} {\bibinfo {author} {\bibfnamefont {H.~R.}\ \bibnamefont
  {Molavian}}, \bibinfo {author} {\bibfnamefont {M.~J.}\ \bibnamefont
  {Gingras}}, \ and\ \bibinfo {author} {\bibfnamefont {B.}~\bibnamefont
  {Canals}},\ }\href@noop {} {\bibfield  {journal} {\bibinfo  {journal}
  {Physical review letters}\ }\textbf {\bibinfo {volume} {98}},\ \bibinfo
  {pages} {157204} (\bibinfo {year} {2007})}\BibitemShut {NoStop}%
\bibitem [{\citenamefont {Gardner}\ \emph {et~al.}(2010)\citenamefont
  {Gardner}, \citenamefont {Gingras},\ and\ \citenamefont
  {Greedan}}]{gardner2010magnetic}%
  \BibitemOpen
  \bibfield  {author} {\bibinfo {author} {\bibfnamefont {J.~S.}\ \bibnamefont
  {Gardner}}, \bibinfo {author} {\bibfnamefont {M.~J.}\ \bibnamefont
  {Gingras}}, \ and\ \bibinfo {author} {\bibfnamefont {J.~E.}\ \bibnamefont
  {Greedan}},\ }\href@noop {} {\bibfield  {journal} {\bibinfo  {journal}
  {Reviews of Modern Physics}\ }\textbf {\bibinfo {volume} {82}},\ \bibinfo
  {pages} {53} (\bibinfo {year} {2010})}\BibitemShut {NoStop}%
\bibitem [{\citenamefont {Thompson}\ \emph {et~al.}(2011)\citenamefont
  {Thompson}, \citenamefont {McClarty}, \citenamefont {R{\o}nnow},
  \citenamefont {Regnault}, \citenamefont {Sorge},\ and\ \citenamefont
  {Gingras}}]{thompson2011rods}%
  \BibitemOpen
  \bibfield  {author} {\bibinfo {author} {\bibfnamefont {J.~D.}\ \bibnamefont
  {Thompson}}, \bibinfo {author} {\bibfnamefont {P.~A.}\ \bibnamefont
  {McClarty}}, \bibinfo {author} {\bibfnamefont {H.~M.}\ \bibnamefont
  {R{\o}nnow}}, \bibinfo {author} {\bibfnamefont {L.~P.}\ \bibnamefont
  {Regnault}}, \bibinfo {author} {\bibfnamefont {A.}~\bibnamefont {Sorge}}, \
  and\ \bibinfo {author} {\bibfnamefont {M.~J.}\ \bibnamefont {Gingras}},\
  }\href@noop {} {\bibfield  {journal} {\bibinfo  {journal} {Physical review
  letters}\ }\textbf {\bibinfo {volume} {106}},\ \bibinfo {pages} {187202}
  (\bibinfo {year} {2011})}\BibitemShut {NoStop}%
\bibitem [{\citenamefont {Ross}\ \emph {et~al.}(2011)\citenamefont {Ross},
  \citenamefont {Savary}, \citenamefont {Gaulin},\ and\ \citenamefont
  {Balents}}]{ross2011quantum}%
  \BibitemOpen
  \bibfield  {author} {\bibinfo {author} {\bibfnamefont {K.~A.}\ \bibnamefont
  {Ross}}, \bibinfo {author} {\bibfnamefont {L.}~\bibnamefont {Savary}},
  \bibinfo {author} {\bibfnamefont {B.~D.}\ \bibnamefont {Gaulin}}, \ and\
  \bibinfo {author} {\bibfnamefont {L.}~\bibnamefont {Balents}},\ }\href@noop
  {} {\bibfield  {journal} {\bibinfo  {journal} {Physical Review X}\ }\textbf
  {\bibinfo {volume} {1}},\ \bibinfo {pages} {021002} (\bibinfo {year}
  {2011})}\BibitemShut {NoStop}%
\bibitem [{\citenamefont {Chang}\ \emph {et~al.}(2012)\citenamefont {Chang},
  \citenamefont {Onoda}, \citenamefont {Su}, \citenamefont {Kao}, \citenamefont
  {Tsuei}, \citenamefont {Yasui}, \citenamefont {Kakurai},\ and\ \citenamefont
  {Lees}}]{chang2012higgs}%
  \BibitemOpen
  \bibfield  {author} {\bibinfo {author} {\bibfnamefont {L.-J.}\ \bibnamefont
  {Chang}}, \bibinfo {author} {\bibfnamefont {S.}~\bibnamefont {Onoda}},
  \bibinfo {author} {\bibfnamefont {Y.}~\bibnamefont {Su}}, \bibinfo {author}
  {\bibfnamefont {Y.-J.}\ \bibnamefont {Kao}}, \bibinfo {author} {\bibfnamefont
  {K.-D.}\ \bibnamefont {Tsuei}}, \bibinfo {author} {\bibfnamefont
  {Y.}~\bibnamefont {Yasui}}, \bibinfo {author} {\bibfnamefont
  {K.}~\bibnamefont {Kakurai}}, \ and\ \bibinfo {author} {\bibfnamefont
  {M.~R.}\ \bibnamefont {Lees}},\ }\href@noop {} {\bibfield  {journal}
  {\bibinfo  {journal} {Nature communications}\ }\textbf {\bibinfo {volume}
  {3}},\ \bibinfo {pages} {992} (\bibinfo {year} {2012})}\BibitemShut {NoStop}%
\bibitem [{\citenamefont {Kimura}\ \emph {et~al.}(2013)\citenamefont {Kimura},
  \citenamefont {Nakatsuji}, \citenamefont {Wen}, \citenamefont {Broholm},
  \citenamefont {Stone}, \citenamefont {Nishibori},\ and\ \citenamefont
  {Sawa}}]{kimura2013quantum}%
  \BibitemOpen
  \bibfield  {author} {\bibinfo {author} {\bibfnamefont {K.}~\bibnamefont
  {Kimura}}, \bibinfo {author} {\bibfnamefont {S.}~\bibnamefont {Nakatsuji}},
  \bibinfo {author} {\bibfnamefont {J.}~\bibnamefont {Wen}}, \bibinfo {author}
  {\bibfnamefont {C.}~\bibnamefont {Broholm}}, \bibinfo {author} {\bibfnamefont
  {M.}~\bibnamefont {Stone}}, \bibinfo {author} {\bibfnamefont
  {E.}~\bibnamefont {Nishibori}}, \ and\ \bibinfo {author} {\bibfnamefont
  {H.}~\bibnamefont {Sawa}},\ }\href@noop {} {\bibfield  {journal} {\bibinfo
  {journal} {Nature communications}\ }\textbf {\bibinfo {volume} {4}},\
  \bibinfo {pages} {1934} (\bibinfo {year} {2013})}\BibitemShut {NoStop}%
\bibitem [{\citenamefont {Pan}\ \emph {et~al.}(2014)\citenamefont {Pan},
  \citenamefont {Kim}, \citenamefont {Ghosh}, \citenamefont {Morris},
  \citenamefont {Ross}, \citenamefont {Kermarrec}, \citenamefont {Gaulin},
  \citenamefont {Koohpayeh}, \citenamefont {Tchernyshyov},\ and\ \citenamefont
  {Armitage}}]{pan2014low}%
  \BibitemOpen
  \bibfield  {author} {\bibinfo {author} {\bibfnamefont {L.}~\bibnamefont
  {Pan}}, \bibinfo {author} {\bibfnamefont {S.~K.}\ \bibnamefont {Kim}},
  \bibinfo {author} {\bibfnamefont {A.}~\bibnamefont {Ghosh}}, \bibinfo
  {author} {\bibfnamefont {C.~M.}\ \bibnamefont {Morris}}, \bibinfo {author}
  {\bibfnamefont {K.~A.}\ \bibnamefont {Ross}}, \bibinfo {author}
  {\bibfnamefont {E.}~\bibnamefont {Kermarrec}}, \bibinfo {author}
  {\bibfnamefont {B.~D.}\ \bibnamefont {Gaulin}}, \bibinfo {author}
  {\bibfnamefont {S.}~\bibnamefont {Koohpayeh}}, \bibinfo {author}
  {\bibfnamefont {O.}~\bibnamefont {Tchernyshyov}}, \ and\ \bibinfo {author}
  {\bibfnamefont {N.}~\bibnamefont {Armitage}},\ }\href@noop {} {\bibfield
  {journal} {\bibinfo  {journal} {Nature communications}\ }\textbf {\bibinfo
  {volume} {5}},\ \bibinfo {pages} {4970} (\bibinfo {year} {2014})}\BibitemShut
  {NoStop}%
\bibitem [{\citenamefont {Pan}\ \emph {et~al.}(2016)\citenamefont {Pan},
  \citenamefont {Laurita}, \citenamefont {Ross}, \citenamefont {Gaulin},\ and\
  \citenamefont {Armitage}}]{pan2016measure}%
  \BibitemOpen
  \bibfield  {author} {\bibinfo {author} {\bibfnamefont {L.}~\bibnamefont
  {Pan}}, \bibinfo {author} {\bibfnamefont {N.}~\bibnamefont {Laurita}},
  \bibinfo {author} {\bibfnamefont {K.~A.}\ \bibnamefont {Ross}}, \bibinfo
  {author} {\bibfnamefont {B.~D.}\ \bibnamefont {Gaulin}}, \ and\ \bibinfo
  {author} {\bibfnamefont {N.}~\bibnamefont {Armitage}},\ }\href@noop {}
  {\bibfield  {journal} {\bibinfo  {journal} {Nature Physics}\ }\textbf
  {\bibinfo {volume} {12}},\ \bibinfo {pages} {361} (\bibinfo {year}
  {2016})}\BibitemShut {NoStop}%
\bibitem [{\citenamefont {Gao}\ \emph {et~al.}(2019)\citenamefont {Gao},
  \citenamefont {Chen}, \citenamefont {Tam}, \citenamefont {Huang},
  \citenamefont {Sasmal}, \citenamefont {Adroja}, \citenamefont {Ye},
  \citenamefont {Cao}, \citenamefont {Sala}, \citenamefont {Stone} \emph
  {et~al.}}]{gao2019experimental}%
  \BibitemOpen
  \bibfield  {author} {\bibinfo {author} {\bibfnamefont {B.}~\bibnamefont
  {Gao}}, \bibinfo {author} {\bibfnamefont {T.}~\bibnamefont {Chen}}, \bibinfo
  {author} {\bibfnamefont {D.~W.}\ \bibnamefont {Tam}}, \bibinfo {author}
  {\bibfnamefont {C.-L.}\ \bibnamefont {Huang}}, \bibinfo {author}
  {\bibfnamefont {K.}~\bibnamefont {Sasmal}}, \bibinfo {author} {\bibfnamefont
  {D.~T.}\ \bibnamefont {Adroja}}, \bibinfo {author} {\bibfnamefont
  {F.}~\bibnamefont {Ye}}, \bibinfo {author} {\bibfnamefont {H.}~\bibnamefont
  {Cao}}, \bibinfo {author} {\bibfnamefont {G.}~\bibnamefont {Sala}}, \bibinfo
  {author} {\bibfnamefont {M.~B.}\ \bibnamefont {Stone}},  \emph {et~al.},\
  }\href@noop {} {\bibfield  {journal} {\bibinfo  {journal} {arXiv preprint
  arXiv:1901.10092}\ } (\bibinfo {year} {2019})}\BibitemShut {NoStop}%
\bibitem [{\citenamefont {Onoda}\ and\ \citenamefont
  {Tanaka}(2010)}]{onoda2010quantum}%
  \BibitemOpen
  \bibfield  {author} {\bibinfo {author} {\bibfnamefont {S.}~\bibnamefont
  {Onoda}}\ and\ \bibinfo {author} {\bibfnamefont {Y.}~\bibnamefont {Tanaka}},\
  }\href@noop {} {\bibfield  {journal} {\bibinfo  {journal} {Physical review
  letters}\ }\textbf {\bibinfo {volume} {105}},\ \bibinfo {pages} {047201}
  (\bibinfo {year} {2010})}\BibitemShut {NoStop}%
\bibitem [{\citenamefont {Onoda}\ and\ \citenamefont
  {Tanaka}(2011)}]{onoda2011quantum}%
  \BibitemOpen
  \bibfield  {author} {\bibinfo {author} {\bibfnamefont {S.}~\bibnamefont
  {Onoda}}\ and\ \bibinfo {author} {\bibfnamefont {Y.}~\bibnamefont {Tanaka}},\
  }\href@noop {} {\bibfield  {journal} {\bibinfo  {journal} {Physical Review
  B}\ }\textbf {\bibinfo {volume} {83}},\ \bibinfo {pages} {094411} (\bibinfo
  {year} {2011})}\BibitemShut {NoStop}%
\bibitem [{\citenamefont {Onoda}(2011)}]{onoda2011effective}%
  \BibitemOpen
  \bibfield  {author} {\bibinfo {author} {\bibfnamefont {S.}~\bibnamefont
  {Onoda}},\ }in\ \href@noop {} {\emph {\bibinfo {booktitle} {Journal of
  Physics: Conference Series}}},\ Vol.\ \bibinfo {volume} {320}\ (\bibinfo
  {organization} {IOP Publishing},\ \bibinfo {year} {2011})\ p.\ \bibinfo
  {pages} {012065}\BibitemShut {NoStop}%
\bibitem [{\citenamefont {Huang}\ \emph {et~al.}(2014)\citenamefont {Huang},
  \citenamefont {Chen},\ and\ \citenamefont {Hermele}}]{huang2014quantum}%
  \BibitemOpen
  \bibfield  {author} {\bibinfo {author} {\bibfnamefont {Y.-P.}\ \bibnamefont
  {Huang}}, \bibinfo {author} {\bibfnamefont {G.}~\bibnamefont {Chen}}, \ and\
  \bibinfo {author} {\bibfnamefont {M.}~\bibnamefont {Hermele}},\ }\href@noop
  {} {\bibfield  {journal} {\bibinfo  {journal} {Physical review letters}\
  }\textbf {\bibinfo {volume} {112}},\ \bibinfo {pages} {167203} (\bibinfo
  {year} {2014})}\BibitemShut {NoStop}%
\bibitem [{\citenamefont {Princep}\ \emph {et~al.}(2015)\citenamefont
  {Princep}, \citenamefont {Walker}, \citenamefont {Adroja}, \citenamefont
  {Prabhakaran},\ and\ \citenamefont {Boothroyd}}]{princep2015crystal}%
  \BibitemOpen
  \bibfield  {author} {\bibinfo {author} {\bibfnamefont {A.}~\bibnamefont
  {Princep}}, \bibinfo {author} {\bibfnamefont {H.}~\bibnamefont {Walker}},
  \bibinfo {author} {\bibfnamefont {D.}~\bibnamefont {Adroja}}, \bibinfo
  {author} {\bibfnamefont {D.}~\bibnamefont {Prabhakaran}}, \ and\ \bibinfo
  {author} {\bibfnamefont {A.}~\bibnamefont {Boothroyd}},\ }\href@noop {}
  {\bibfield  {journal} {\bibinfo  {journal} {Physical Review B}\ }\textbf
  {\bibinfo {volume} {91}},\ \bibinfo {pages} {224430} (\bibinfo {year}
  {2015})}\BibitemShut {NoStop}%
\bibitem [{\citenamefont {Ruminy}\ \emph {et~al.}(2016)\citenamefont {Ruminy},
  \citenamefont {Pomjakushina}, \citenamefont {Iida}, \citenamefont {Kamazawa},
  \citenamefont {Adroja}, \citenamefont {Stuhr},\ and\ \citenamefont
  {Fennell}}]{ruminy2016crystal}%
  \BibitemOpen
  \bibfield  {author} {\bibinfo {author} {\bibfnamefont {M.}~\bibnamefont
  {Ruminy}}, \bibinfo {author} {\bibfnamefont {E.}~\bibnamefont
  {Pomjakushina}}, \bibinfo {author} {\bibfnamefont {K.}~\bibnamefont {Iida}},
  \bibinfo {author} {\bibfnamefont {K.}~\bibnamefont {Kamazawa}}, \bibinfo
  {author} {\bibfnamefont {D.}~\bibnamefont {Adroja}}, \bibinfo {author}
  {\bibfnamefont {U.}~\bibnamefont {Stuhr}}, \ and\ \bibinfo {author}
  {\bibfnamefont {T.}~\bibnamefont {Fennell}},\ }\href@noop {} {\bibfield
  {journal} {\bibinfo  {journal} {Physical Review B}\ }\textbf {\bibinfo
  {volume} {94}},\ \bibinfo {pages} {024430} (\bibinfo {year}
  {2016})}\BibitemShut {NoStop}%
\bibitem [{\citenamefont {Ramirez}\ \emph {et~al.}(1999)\citenamefont
  {Ramirez}, \citenamefont {Hayashi}, \citenamefont {Cava}, \citenamefont
  {Siddharthan},\ and\ \citenamefont {Shastry}}]{ramirez1999zero}%
  \BibitemOpen
  \bibfield  {author} {\bibinfo {author} {\bibfnamefont {A.~P.}\ \bibnamefont
  {Ramirez}}, \bibinfo {author} {\bibfnamefont {A.}~\bibnamefont {Hayashi}},
  \bibinfo {author} {\bibfnamefont {R.~J.}\ \bibnamefont {Cava}}, \bibinfo
  {author} {\bibfnamefont {R.}~\bibnamefont {Siddharthan}}, \ and\ \bibinfo
  {author} {\bibfnamefont {B.}~\bibnamefont {Shastry}},\ }\href@noop {}
  {\bibfield  {journal} {\bibinfo  {journal} {Nature}\ }\textbf {\bibinfo
  {volume} {399}},\ \bibinfo {pages} {333} (\bibinfo {year}
  {1999})}\BibitemShut {NoStop}%
\bibitem [{\citenamefont {Moessner}(2001)}]{moessner2001magnets}%
  \BibitemOpen
  \bibfield  {author} {\bibinfo {author} {\bibfnamefont {R.}~\bibnamefont
  {Moessner}},\ }\href@noop {} {\bibfield  {journal} {\bibinfo  {journal}
  {Canadian journal of physics}\ }\textbf {\bibinfo {volume} {79}},\ \bibinfo
  {pages} {1283} (\bibinfo {year} {2001})}\BibitemShut {NoStop}%
\bibitem [{\citenamefont {Moessner}\ and\ \citenamefont
  {Ramirez}(2006)}]{moessner2006geometrical}%
  \BibitemOpen
  \bibfield  {author} {\bibinfo {author} {\bibfnamefont {R.}~\bibnamefont
  {Moessner}}\ and\ \bibinfo {author} {\bibfnamefont {A.~P.}\ \bibnamefont
  {Ramirez}},\ }\href@noop {} {\bibfield  {journal} {\bibinfo  {journal} {Phys.
  Today}\ }\textbf {\bibinfo {volume} {59}},\ \bibinfo {pages} {24} (\bibinfo
  {year} {2006})}\BibitemShut {NoStop}%
\bibitem [{\citenamefont {Castelnovo}\ \emph {et~al.}(2008)\citenamefont
  {Castelnovo}, \citenamefont {Moessner},\ and\ \citenamefont
  {Sondhi}}]{castelnovo2008magnetic}%
  \BibitemOpen
  \bibfield  {author} {\bibinfo {author} {\bibfnamefont {C.}~\bibnamefont
  {Castelnovo}}, \bibinfo {author} {\bibfnamefont {R.}~\bibnamefont
  {Moessner}}, \ and\ \bibinfo {author} {\bibfnamefont {S.~L.}\ \bibnamefont
  {Sondhi}},\ }\href@noop {} {\bibfield  {journal} {\bibinfo  {journal}
  {Nature}\ }\textbf {\bibinfo {volume} {451}},\ \bibinfo {pages} {42}
  (\bibinfo {year} {2008})}\BibitemShut {NoStop}%
\bibitem [{\citenamefont {Moessner}\ and\ \citenamefont
  {Sondhi}(2003)}]{moessner2003three}%
  \BibitemOpen
  \bibfield  {author} {\bibinfo {author} {\bibfnamefont {R.}~\bibnamefont
  {Moessner}}\ and\ \bibinfo {author} {\bibfnamefont {S.~L.}\ \bibnamefont
  {Sondhi}},\ }\href@noop {} {\bibfield  {journal} {\bibinfo  {journal}
  {Physical Review B}\ }\textbf {\bibinfo {volume} {68}},\ \bibinfo {pages}
  {184512} (\bibinfo {year} {2003})}\BibitemShut {NoStop}%
\bibitem [{\citenamefont {Hermele}\ \emph {et~al.}(2004)\citenamefont
  {Hermele}, \citenamefont {Fisher},\ and\ \citenamefont
  {Balents}}]{hermele2004pyrochlore}%
  \BibitemOpen
  \bibfield  {author} {\bibinfo {author} {\bibfnamefont {M.}~\bibnamefont
  {Hermele}}, \bibinfo {author} {\bibfnamefont {M.~P.}\ \bibnamefont {Fisher}},
  \ and\ \bibinfo {author} {\bibfnamefont {L.}~\bibnamefont {Balents}},\
  }\href@noop {} {\bibfield  {journal} {\bibinfo  {journal} {Physical Review
  B}\ }\textbf {\bibinfo {volume} {69}},\ \bibinfo {pages} {064404} (\bibinfo
  {year} {2004})}\BibitemShut {NoStop}%
\bibitem [{\citenamefont {Sikora}\ \emph {et~al.}(2009)\citenamefont {Sikora},
  \citenamefont {Pollmann}, \citenamefont {Shannon}, \citenamefont {Penc},\
  and\ \citenamefont {Fulde}}]{sikora2009quantum}%
  \BibitemOpen
  \bibfield  {author} {\bibinfo {author} {\bibfnamefont {O.}~\bibnamefont
  {Sikora}}, \bibinfo {author} {\bibfnamefont {F.}~\bibnamefont {Pollmann}},
  \bibinfo {author} {\bibfnamefont {N.}~\bibnamefont {Shannon}}, \bibinfo
  {author} {\bibfnamefont {K.}~\bibnamefont {Penc}}, \ and\ \bibinfo {author}
  {\bibfnamefont {P.}~\bibnamefont {Fulde}},\ }\href@noop {} {\bibfield
  {journal} {\bibinfo  {journal} {Physical review letters}\ }\textbf {\bibinfo
  {volume} {103}},\ \bibinfo {pages} {247001} (\bibinfo {year}
  {2009})}\BibitemShut {NoStop}%
\bibitem [{\citenamefont {Savary}\ and\ \citenamefont
  {Balents}(2012)}]{savary2012coulombic}%
  \BibitemOpen
  \bibfield  {author} {\bibinfo {author} {\bibfnamefont {L.}~\bibnamefont
  {Savary}}\ and\ \bibinfo {author} {\bibfnamefont {L.}~\bibnamefont
  {Balents}},\ }\href@noop {} {\bibfield  {journal} {\bibinfo  {journal}
  {Physical review letters}\ }\textbf {\bibinfo {volume} {108}},\ \bibinfo
  {pages} {037202} (\bibinfo {year} {2012})}\BibitemShut {NoStop}%
\bibitem [{\citenamefont {Lee}\ \emph {et~al.}(2012)\citenamefont {Lee},
  \citenamefont {Onoda},\ and\ \citenamefont {Balents}}]{lee2012generic}%
  \BibitemOpen
  \bibfield  {author} {\bibinfo {author} {\bibfnamefont {S.}~\bibnamefont
  {Lee}}, \bibinfo {author} {\bibfnamefont {S.}~\bibnamefont {Onoda}}, \ and\
  \bibinfo {author} {\bibfnamefont {L.}~\bibnamefont {Balents}},\ }\href@noop
  {} {\bibfield  {journal} {\bibinfo  {journal} {Physical Review B}\ }\textbf
  {\bibinfo {volume} {86}},\ \bibinfo {pages} {104412} (\bibinfo {year}
  {2012})}\BibitemShut {NoStop}%
\bibitem [{\citenamefont {Shannon}\ \emph {et~al.}(2012)\citenamefont
  {Shannon}, \citenamefont {Sikora}, \citenamefont {Pollmann}, \citenamefont
  {Penc},\ and\ \citenamefont {Fulde}}]{shannon2012quantum}%
  \BibitemOpen
  \bibfield  {author} {\bibinfo {author} {\bibfnamefont {N.}~\bibnamefont
  {Shannon}}, \bibinfo {author} {\bibfnamefont {O.}~\bibnamefont {Sikora}},
  \bibinfo {author} {\bibfnamefont {F.}~\bibnamefont {Pollmann}}, \bibinfo
  {author} {\bibfnamefont {K.}~\bibnamefont {Penc}}, \ and\ \bibinfo {author}
  {\bibfnamefont {P.}~\bibnamefont {Fulde}},\ }\href@noop {} {\bibfield
  {journal} {\bibinfo  {journal} {Physical review letters}\ }\textbf {\bibinfo
  {volume} {108}},\ \bibinfo {pages} {067204} (\bibinfo {year}
  {2012})}\BibitemShut {NoStop}%
\bibitem [{\citenamefont {Lantagne-Hurtubise}\ \emph
  {et~al.}(2017)\citenamefont {Lantagne-Hurtubise}, \citenamefont
  {Bhattacharjee},\ and\ \citenamefont {Moessner}}]{lantagne2017electric}%
  \BibitemOpen
  \bibfield  {author} {\bibinfo {author} {\bibfnamefont {{\'E}.}~\bibnamefont
  {Lantagne-Hurtubise}}, \bibinfo {author} {\bibfnamefont {S.}~\bibnamefont
  {Bhattacharjee}}, \ and\ \bibinfo {author} {\bibfnamefont {R.}~\bibnamefont
  {Moessner}},\ }\href@noop {} {\bibfield  {journal} {\bibinfo  {journal}
  {Physical Review B}\ }\textbf {\bibinfo {volume} {96}},\ \bibinfo {pages}
  {125145} (\bibinfo {year} {2017})}\BibitemShut {NoStop}%
\bibitem [{\citenamefont {Wen}(2002)}]{wen2002quantum}%
  \BibitemOpen
  \bibfield  {author} {\bibinfo {author} {\bibfnamefont {X.-G.}\ \bibnamefont
  {Wen}},\ }\href@noop {} {\bibfield  {journal} {\bibinfo  {journal} {Physical
  Review B}\ }\textbf {\bibinfo {volume} {65}},\ \bibinfo {pages} {165113}
  (\bibinfo {year} {2002})}\BibitemShut {NoStop}%
\bibitem [{\citenamefont {Chen}(2017)}]{chen2017spectral}%
  \BibitemOpen
  \bibfield  {author} {\bibinfo {author} {\bibfnamefont {G.}~\bibnamefont
  {Chen}},\ }\href@noop {} {\bibfield  {journal} {\bibinfo  {journal} {Physical
  Review B}\ }\textbf {\bibinfo {volume} {96}},\ \bibinfo {pages} {085136}
  (\bibinfo {year} {2017})}\BibitemShut {NoStop}%
\bibitem [{\citenamefont {Hirschberger}\ \emph
  {et~al.}(2015{\natexlab{a}})\citenamefont {Hirschberger}, \citenamefont
  {Krizan}, \citenamefont {Cava},\ and\ \citenamefont
  {Ong}}]{hirschberger2015large}%
  \BibitemOpen
  \bibfield  {author} {\bibinfo {author} {\bibfnamefont {M.}~\bibnamefont
  {Hirschberger}}, \bibinfo {author} {\bibfnamefont {J.~W.}\ \bibnamefont
  {Krizan}}, \bibinfo {author} {\bibfnamefont {R.~J.}\ \bibnamefont {Cava}}, \
  and\ \bibinfo {author} {\bibfnamefont {N.~P.}\ \bibnamefont {Ong}},\
  }\href@noop {} {\bibfield  {journal} {\bibinfo  {journal} {Science}\ }\textbf
  {\bibinfo {volume} {348}},\ \bibinfo {pages} {106} (\bibinfo {year}
  {2015}{\natexlab{a}})}\BibitemShut {NoStop}%
\bibitem [{\citenamefont {Gao}\ and\ \citenamefont
  {Chen}(2019)}]{gao2019topological}%
  \BibitemOpen
  \bibfield  {author} {\bibinfo {author} {\bibfnamefont {Y.~H.}\ \bibnamefont
  {Gao}}\ and\ \bibinfo {author} {\bibfnamefont {G.}~\bibnamefont {Chen}},\
  }\href@noop {} {\bibfield  {journal} {\bibinfo  {journal} {arXiv preprint
  arXiv:1901.01522}\ } (\bibinfo {year} {2019})}\BibitemShut {NoStop}%
\bibitem [{\citenamefont {Zhang}\ \emph {et~al.}(2019)\citenamefont {Zhang},
  \citenamefont {Gao}, \citenamefont {Liu},\ and\ \citenamefont
  {Chen}}]{zhang2019topological}%
  \BibitemOpen
  \bibfield  {author} {\bibinfo {author} {\bibfnamefont {X.-T.}\ \bibnamefont
  {Zhang}}, \bibinfo {author} {\bibfnamefont {Y.~H.}\ \bibnamefont {Gao}},
  \bibinfo {author} {\bibfnamefont {C.}~\bibnamefont {Liu}}, \ and\ \bibinfo
  {author} {\bibfnamefont {G.}~\bibnamefont {Chen}},\ }\href@noop {} {\bibfield
   {journal} {\bibinfo  {journal} {arXiv preprint arXiv:1904.08865}\ }
  (\bibinfo {year} {2019})}\BibitemShut {NoStop}%
\bibitem [{\citenamefont {Motrunich}\ and\ \citenamefont
  {Senthil}(2002)}]{motrunich2002exotic}%
  \BibitemOpen
  \bibfield  {author} {\bibinfo {author} {\bibfnamefont {O.}~\bibnamefont
  {Motrunich}}\ and\ \bibinfo {author} {\bibfnamefont {T.}~\bibnamefont
  {Senthil}},\ }\href@noop {} {\bibfield  {journal} {\bibinfo  {journal}
  {Physical review letters}\ }\textbf {\bibinfo {volume} {89}},\ \bibinfo
  {pages} {277004} (\bibinfo {year} {2002})}\BibitemShut {NoStop}%
\bibitem [{\citenamefont {Motrunich}\ and\ \citenamefont
  {Senthil}(2005)}]{motrunich2005origin}%
  \BibitemOpen
  \bibfield  {author} {\bibinfo {author} {\bibfnamefont {O.}~\bibnamefont
  {Motrunich}}\ and\ \bibinfo {author} {\bibfnamefont {T.}~\bibnamefont
  {Senthil}},\ }\href@noop {} {\bibfield  {journal} {\bibinfo  {journal}
  {Physical Review B}\ }\textbf {\bibinfo {volume} {71}},\ \bibinfo {pages}
  {125102} (\bibinfo {year} {2005})}\BibitemShut {NoStop}%
\bibitem [{\citenamefont {Jaubert}\ and\ \citenamefont
  {Holdsworth}(2011)}]{jaubert2011magnetic}%
  \BibitemOpen
  \bibfield  {author} {\bibinfo {author} {\bibfnamefont {L.~D.}\ \bibnamefont
  {Jaubert}}\ and\ \bibinfo {author} {\bibfnamefont {P.~C.}\ \bibnamefont
  {Holdsworth}},\ }\href@noop {} {\bibfield  {journal} {\bibinfo  {journal}
  {Journal of Physics: Condensed Matter}\ }\textbf {\bibinfo {volume} {23}},\
  \bibinfo {pages} {164222} (\bibinfo {year} {2011})}\BibitemShut {NoStop}%
\bibitem [{\citenamefont {Essin}\ and\ \citenamefont
  {Hermele}(2013)}]{essin2013classifying}%
  \BibitemOpen
  \bibfield  {author} {\bibinfo {author} {\bibfnamefont {A.~M.}\ \bibnamefont
  {Essin}}\ and\ \bibinfo {author} {\bibfnamefont {M.}~\bibnamefont
  {Hermele}},\ }\href@noop {} {\bibfield  {journal} {\bibinfo  {journal}
  {Physical Review B}\ }\textbf {\bibinfo {volume} {87}},\ \bibinfo {pages}
  {104406} (\bibinfo {year} {2013})}\BibitemShut {NoStop}%
\bibitem [{\citenamefont {Isakov}\ \emph {et~al.}(2004)\citenamefont {Isakov},
  \citenamefont {Gregor}, \citenamefont {Moessner},\ and\ \citenamefont
  {Sondhi}}]{isakov2004dipolar}%
  \BibitemOpen
  \bibfield  {author} {\bibinfo {author} {\bibfnamefont {S.}~\bibnamefont
  {Isakov}}, \bibinfo {author} {\bibfnamefont {K.}~\bibnamefont {Gregor}},
  \bibinfo {author} {\bibfnamefont {R.}~\bibnamefont {Moessner}}, \ and\
  \bibinfo {author} {\bibfnamefont {S.~L.}\ \bibnamefont {Sondhi}},\
  }\href@noop {} {\bibfield  {journal} {\bibinfo  {journal} {Physical review
  letters}\ }\textbf {\bibinfo {volume} {93}},\ \bibinfo {pages} {167204}
  (\bibinfo {year} {2004})}\BibitemShut {NoStop}%
\bibitem [{\citenamefont {Fennell}\ \emph {et~al.}(2009)\citenamefont
  {Fennell}, \citenamefont {Deen}, \citenamefont {Wildes}, \citenamefont
  {Schmalzl}, \citenamefont {Prabhakaran}, \citenamefont {Boothroyd},
  \citenamefont {Aldus}, \citenamefont {McMorrow},\ and\ \citenamefont
  {Bramwell}}]{fennell2009magnetic}%
  \BibitemOpen
  \bibfield  {author} {\bibinfo {author} {\bibfnamefont {T.}~\bibnamefont
  {Fennell}}, \bibinfo {author} {\bibfnamefont {P.}~\bibnamefont {Deen}},
  \bibinfo {author} {\bibfnamefont {A.}~\bibnamefont {Wildes}}, \bibinfo
  {author} {\bibfnamefont {K.}~\bibnamefont {Schmalzl}}, \bibinfo {author}
  {\bibfnamefont {D.}~\bibnamefont {Prabhakaran}}, \bibinfo {author}
  {\bibfnamefont {A.}~\bibnamefont {Boothroyd}}, \bibinfo {author}
  {\bibfnamefont {R.}~\bibnamefont {Aldus}}, \bibinfo {author} {\bibfnamefont
  {D.}~\bibnamefont {McMorrow}}, \ and\ \bibinfo {author} {\bibfnamefont
  {S.}~\bibnamefont {Bramwell}},\ }\href@noop {} {\bibfield  {journal}
  {\bibinfo  {journal} {Science}\ }\textbf {\bibinfo {volume} {326}},\ \bibinfo
  {pages} {415} (\bibinfo {year} {2009})}\BibitemShut {NoStop}%
\bibitem [{\citenamefont {Castelnovo}\ \emph {et~al.}(2012)\citenamefont
  {Castelnovo}, \citenamefont {Moessner},\ and\ \citenamefont
  {Sondhi}}]{castelnovo2012spin}%
  \BibitemOpen
  \bibfield  {author} {\bibinfo {author} {\bibfnamefont {C.}~\bibnamefont
  {Castelnovo}}, \bibinfo {author} {\bibfnamefont {R.}~\bibnamefont
  {Moessner}}, \ and\ \bibinfo {author} {\bibfnamefont {S.~L.}\ \bibnamefont
  {Sondhi}},\ }\href@noop {} {\bibfield  {journal} {\bibinfo  {journal} {Annu.
  Rev. Condens. Matter Phys.}\ }\textbf {\bibinfo {volume} {3}},\ \bibinfo
  {pages} {35} (\bibinfo {year} {2012})}\BibitemShut {NoStop}%
\bibitem [{\citenamefont {Polyakov}(1975)}]{polyakov1975interaction}%
  \BibitemOpen
  \bibfield  {author} {\bibinfo {author} {\bibfnamefont {A.~M.}\ \bibnamefont
  {Polyakov}},\ }\href@noop {} {\bibfield  {journal} {\bibinfo  {journal}
  {Physics Letters B}\ }\textbf {\bibinfo {volume} {59}},\ \bibinfo {pages}
  {79} (\bibinfo {year} {1975})}\BibitemShut {NoStop}%
\bibitem [{\citenamefont {Hofstadter}(1976)}]{hofstadter1976energy}%
  \BibitemOpen
  \bibfield  {author} {\bibinfo {author} {\bibfnamefont {D.~R.}\ \bibnamefont
  {Hofstadter}},\ }\href@noop {} {\bibfield  {journal} {\bibinfo  {journal}
  {Physical review B}\ }\textbf {\bibinfo {volume} {14}},\ \bibinfo {pages}
  {2239} (\bibinfo {year} {1976})}\BibitemShut {NoStop}%
\bibitem [{\citenamefont {Bernevig}\ and\ \citenamefont
  {Hughes}(2013)}]{bernevig2013topological}%
  \BibitemOpen
  \bibfield  {author} {\bibinfo {author} {\bibfnamefont {B.~A.}\ \bibnamefont
  {Bernevig}}\ and\ \bibinfo {author} {\bibfnamefont {T.~L.}\ \bibnamefont
  {Hughes}},\ }\href@noop {} {\emph {\bibinfo {title} {Topological insulators
  and topological superconductors}}}\ (\bibinfo  {publisher} {Princeton
  university press},\ \bibinfo {year} {2013})\BibitemShut {NoStop}%
\bibitem [{\citenamefont {Katsura}\ \emph {et~al.}(2010)\citenamefont
  {Katsura}, \citenamefont {Nagaosa},\ and\ \citenamefont
  {Lee}}]{katsura2010theory}%
  \BibitemOpen
  \bibfield  {author} {\bibinfo {author} {\bibfnamefont {H.}~\bibnamefont
  {Katsura}}, \bibinfo {author} {\bibfnamefont {N.}~\bibnamefont {Nagaosa}}, \
  and\ \bibinfo {author} {\bibfnamefont {P.~A.}\ \bibnamefont {Lee}},\
  }\href@noop {} {\bibfield  {journal} {\bibinfo  {journal} {Physical review
  letters}\ }\textbf {\bibinfo {volume} {104}},\ \bibinfo {pages} {066403}
  (\bibinfo {year} {2010})}\BibitemShut {NoStop}%
\bibitem [{\citenamefont {Murakami}\ and\ \citenamefont
  {Okamoto}(2016)}]{murakami2016thermal}%
  \BibitemOpen
  \bibfield  {author} {\bibinfo {author} {\bibfnamefont {S.}~\bibnamefont
  {Murakami}}\ and\ \bibinfo {author} {\bibfnamefont {A.}~\bibnamefont
  {Okamoto}},\ }\href@noop {} {\bibfield  {journal} {\bibinfo  {journal}
  {Journal of the Physical Society of Japan}\ }\textbf {\bibinfo {volume}
  {86}},\ \bibinfo {pages} {011010} (\bibinfo {year} {2016})}\BibitemShut
  {NoStop}%
\bibitem [{\citenamefont {Luttinger}(1964)}]{luttinger1964theory}%
  \BibitemOpen
  \bibfield  {author} {\bibinfo {author} {\bibfnamefont {J.}~\bibnamefont
  {Luttinger}},\ }\href@noop {} {\bibfield  {journal} {\bibinfo  {journal}
  {Physical Review}\ }\textbf {\bibinfo {volume} {135}},\ \bibinfo {pages}
  {A1505} (\bibinfo {year} {1964})}\BibitemShut {NoStop}%
\bibitem [{\citenamefont {Smrcka}\ and\ \citenamefont
  {Streda}(1977)}]{smrcka1977transport}%
  \BibitemOpen
  \bibfield  {author} {\bibinfo {author} {\bibfnamefont {L.}~\bibnamefont
  {Smrcka}}\ and\ \bibinfo {author} {\bibfnamefont {P.}~\bibnamefont
  {Streda}},\ }\href@noop {} {\bibfield  {journal} {\bibinfo  {journal}
  {Journal of Physics C: Solid State Physics}\ }\textbf {\bibinfo {volume}
  {10}},\ \bibinfo {pages} {2153} (\bibinfo {year} {1977})}\BibitemShut
  {NoStop}%
\bibitem [{\citenamefont {Matsumoto}\ and\ \citenamefont
  {Murakami}(2011)}]{matsumoto2011theoretical}%
  \BibitemOpen
  \bibfield  {author} {\bibinfo {author} {\bibfnamefont {R.}~\bibnamefont
  {Matsumoto}}\ and\ \bibinfo {author} {\bibfnamefont {S.}~\bibnamefont
  {Murakami}},\ }\href@noop {} {\bibfield  {journal} {\bibinfo  {journal}
  {Physical review letters}\ }\textbf {\bibinfo {volume} {106}},\ \bibinfo
  {pages} {197202} (\bibinfo {year} {2011})}\BibitemShut {NoStop}%
\bibitem [{\citenamefont {Matsumoto}\ \emph {et~al.}(2014)\citenamefont
  {Matsumoto}, \citenamefont {Shindou},\ and\ \citenamefont
  {Murakami}}]{matsumoto2014thermal}%
  \BibitemOpen
  \bibfield  {author} {\bibinfo {author} {\bibfnamefont {R.}~\bibnamefont
  {Matsumoto}}, \bibinfo {author} {\bibfnamefont {R.}~\bibnamefont {Shindou}},
  \ and\ \bibinfo {author} {\bibfnamefont {S.}~\bibnamefont {Murakami}},\
  }\href@noop {} {\bibfield  {journal} {\bibinfo  {journal} {Physical Review
  B}\ }\textbf {\bibinfo {volume} {89}},\ \bibinfo {pages} {054420} (\bibinfo
  {year} {2014})}\BibitemShut {NoStop}%
\bibitem [{\citenamefont {Hirschberger}\ \emph
  {et~al.}(2015{\natexlab{b}})\citenamefont {Hirschberger}, \citenamefont
  {Chisnell}, \citenamefont {Lee},\ and\ \citenamefont
  {Ong}}]{hirschberger2015thermal}%
  \BibitemOpen
  \bibfield  {author} {\bibinfo {author} {\bibfnamefont {M.}~\bibnamefont
  {Hirschberger}}, \bibinfo {author} {\bibfnamefont {R.}~\bibnamefont
  {Chisnell}}, \bibinfo {author} {\bibfnamefont {Y.~S.}\ \bibnamefont {Lee}}, \
  and\ \bibinfo {author} {\bibfnamefont {N.~P.}\ \bibnamefont {Ong}},\
  }\href@noop {} {\bibfield  {journal} {\bibinfo  {journal} {Physical review
  letters}\ }\textbf {\bibinfo {volume} {115}},\ \bibinfo {pages} {106603}
  (\bibinfo {year} {2015}{\natexlab{b}})}\BibitemShut {NoStop}%
\bibitem [{\citenamefont {Kasahara}\ \emph {et~al.}(2018)\citenamefont
  {Kasahara}, \citenamefont {Ohnishi}, \citenamefont {Mizukami}, \citenamefont
  {Tanaka}, \citenamefont {Ma}, \citenamefont {Sugii}, \citenamefont {Kurita},
  \citenamefont {Tanaka}, \citenamefont {Nasu}, \citenamefont {Motome} \emph
  {et~al.}}]{kasahara2018majorana}%
  \BibitemOpen
  \bibfield  {author} {\bibinfo {author} {\bibfnamefont {Y.}~\bibnamefont
  {Kasahara}}, \bibinfo {author} {\bibfnamefont {T.}~\bibnamefont {Ohnishi}},
  \bibinfo {author} {\bibfnamefont {Y.}~\bibnamefont {Mizukami}}, \bibinfo
  {author} {\bibfnamefont {O.}~\bibnamefont {Tanaka}}, \bibinfo {author}
  {\bibfnamefont {S.}~\bibnamefont {Ma}}, \bibinfo {author} {\bibfnamefont
  {K.}~\bibnamefont {Sugii}}, \bibinfo {author} {\bibfnamefont
  {N.}~\bibnamefont {Kurita}}, \bibinfo {author} {\bibfnamefont
  {H.}~\bibnamefont {Tanaka}}, \bibinfo {author} {\bibfnamefont
  {J.}~\bibnamefont {Nasu}}, \bibinfo {author} {\bibfnamefont {Y.}~\bibnamefont
  {Motome}},  \emph {et~al.},\ }\href@noop {} {\bibfield  {journal} {\bibinfo
  {journal} {Nature}\ }\textbf {\bibinfo {volume} {559}},\ \bibinfo {pages}
  {227} (\bibinfo {year} {2018})}\BibitemShut {NoStop}%
\bibitem [{\citenamefont {Yu}\ \emph {et~al.}(2018)\citenamefont {Yu},
  \citenamefont {Xu}, \citenamefont {Ran}, \citenamefont {Ni}, \citenamefont
  {Huang}, \citenamefont {Wang}, \citenamefont {Wen},\ and\ \citenamefont
  {Li}}]{yu2018ultralow}%
  \BibitemOpen
  \bibfield  {author} {\bibinfo {author} {\bibfnamefont {Y.}~\bibnamefont
  {Yu}}, \bibinfo {author} {\bibfnamefont {Y.}~\bibnamefont {Xu}}, \bibinfo
  {author} {\bibfnamefont {K.}~\bibnamefont {Ran}}, \bibinfo {author}
  {\bibfnamefont {J.}~\bibnamefont {Ni}}, \bibinfo {author} {\bibfnamefont
  {Y.}~\bibnamefont {Huang}}, \bibinfo {author} {\bibfnamefont
  {J.}~\bibnamefont {Wang}}, \bibinfo {author} {\bibfnamefont {J.}~\bibnamefont
  {Wen}}, \ and\ \bibinfo {author} {\bibfnamefont {S.}~\bibnamefont {Li}},\
  }\href@noop {} {\bibfield  {journal} {\bibinfo  {journal} {Physical review
  letters}\ }\textbf {\bibinfo {volume} {120}},\ \bibinfo {pages} {067202}
  (\bibinfo {year} {2018})}\BibitemShut {NoStop}%
\bibitem [{\citenamefont {Hirschberger}\ \emph {et~al.}(2019)\citenamefont
  {Hirschberger}, \citenamefont {Czajka}, \citenamefont {Koohpayeh},
  \citenamefont {Wang},\ and\ \citenamefont {Ong}}]{hirschberger2019enhanced}%
  \BibitemOpen
  \bibfield  {author} {\bibinfo {author} {\bibfnamefont {M.}~\bibnamefont
  {Hirschberger}}, \bibinfo {author} {\bibfnamefont {P.}~\bibnamefont
  {Czajka}}, \bibinfo {author} {\bibfnamefont {S.}~\bibnamefont {Koohpayeh}},
  \bibinfo {author} {\bibfnamefont {W.}~\bibnamefont {Wang}}, \ and\ \bibinfo
  {author} {\bibfnamefont {N.~P.}\ \bibnamefont {Ong}},\ }\href@noop {}
  {\bibfield  {journal} {\bibinfo  {journal} {arXiv preprint arXiv:1903.00595}\
  } (\bibinfo {year} {2019})}\BibitemShut {NoStop}%
\bibitem [{\citenamefont {Hao}\ \emph {et~al.}(2014)\citenamefont {Hao},
  \citenamefont {Day},\ and\ \citenamefont {Gingras}}]{hao2014bosonic}%
  \BibitemOpen
  \bibfield  {author} {\bibinfo {author} {\bibfnamefont {Z.}~\bibnamefont
  {Hao}}, \bibinfo {author} {\bibfnamefont {A.~G.}\ \bibnamefont {Day}}, \ and\
  \bibinfo {author} {\bibfnamefont {M.~J.}\ \bibnamefont {Gingras}},\
  }\href@noop {} {\bibfield  {journal} {\bibinfo  {journal} {Physical Review
  B}\ }\textbf {\bibinfo {volume} {90}},\ \bibinfo {pages} {214430} (\bibinfo
  {year} {2014})}\BibitemShut {NoStop}%
\end{thebibliography}%
\newpage
\thispagestyle{empty}
\mbox{}
\pagebreak
\newpage
\onecolumngrid
\begin{center}
  \textbf{\large Magnetic field and thermal Hall effect in a pyrochlore U(1) quantum spin liquid\\Supplementary Information}\\[.2cm]
  
  Hyeok-Jun Yang,$^{1}$ Hee Seung Kim,$^{1}$ and SungBin Lee$^1$\\[.1cm]
  {\itshape ${}^1$Department of Physics, Korea Advanced Institute of Science and Technology, Daejeon, 34141, Korea\\
}
(Dated: \today)\\[1cm]
\end{center}
\twocolumngrid

\label{sec: Local axis}
\section{I. Local axis coordinate on the pyrochlore lattice}
The local coordinate axis in which the pseudospin $\textbf{S}_i$ in Eqs. (\ref{eq:Ham}), (\ref{eq:Zeeman}) resides are \cite{ross2011quantum}
\bea
\hat{\textbf{e}}_0&=&(1,1,1)/\sqrt{3},\quad\quad \hat{\textbf{a}}_0=(-2,1,1)/\sqrt{6} \nonumber\\
\hat{\textbf{e}}_1&=&(1,-1,-1)/\sqrt{3},\;\; \hat{\textbf{a}}_1=(-2,-1,-1)/\sqrt{6} \nonumber\\
\hat{\textbf{e}}_2&=&(-1,1,-1)/\sqrt{3},\;\; \hat{\textbf{a}}_2=(2,1,-1)/\sqrt{6} \nonumber\\
\hat{\textbf{e}}_3&=&(-1,-1,1)/\sqrt{3},\;\; \hat{\textbf{a}}_3=(2,-1,1)/\sqrt{6}
\label{eq:Localbase}
\eea
and $\hat{\textbf{b}}_i=\hat{\textbf{e}}_i \times \hat{\textbf{a}}_i$ on each $i^{\text{th}}$ sublattice. When the B-field is applied along [110]-direction ([111]-direction), the in-plane polar angles are
\bea
\text{[110]-direction}:& \nonumber\\
\phi_0 = -\frac{2\pi}{3}&,&\phi_1 = \frac{5\pi}{6}, \; \phi_2 = -\frac{\pi}{6},\; \phi_3 = \frac{\pi}{3} 
\nonumber\\
\text{[111]-direction}:& \nonumber\\
\phi_0 = 0,\; \phi_1 &=& \pi, \; \phi_2 = -\frac{\pi}{3},\; \phi_3 = \frac{\pi}{3} 
\label{eq:Polar}
\eea
so that the spin exchange in Eq. (\ref{eq:Zeeman}) is rewritten as
\bea
H^{\pm}_{Z}=-\frac{1}{2}\sum_i{h^{\perp}_i}(S_i^+e^{-i\phi_i}+S_i^-e^{i\phi_i})
\label{eq:PolarZeeman}
\eea
where $S_{i}^{\pm}=S^x\pm iS^y$. The in-plane component of B-field is $h^{\perp}_0=h^{\perp}_3=h/\sqrt{3}$, $h^{\perp}_1=h^{\perp}_2=h$ for [110]-direction and $h^{\perp}_0=0$, $h^{\perp}_1=h^{\perp}_2=h^{\perp}_3=2\sqrt{2}h/3$ for [111]-direction.

Generically, the symmetry allowed exchange interactions other than Eq. (\ref{eq:Ham}) consist of \cite{ross2011quantum, savary2012coulombic, lee2012generic}
\bea
&H^{\pm\pm}&+H^{\pm z}
\nonumber\\
&&=\frac{J_{\pm\pm}}{2}\sum_{\langle i,j\rangle}(\gamma_{ij}S_i^+S_j^++\gamma_{ij}^*S_i^-S_j^-)
\nonumber\\
&&+\frac{J_{\pm z}}{2}\sum_{\langle i,j\rangle}(S_i^{z}(\zeta_{ij}S_j^++\zeta_{ij}^*S_j^-)+ i \leftrightarrow j)\quad\quad
\label{eq:OtherEx}
\eea
where $\gamma_{ij}$ and $\zeta_{ij}$ are $4 \times 4$ unimodular complex matrices.
\begin{gather}
\zeta=-\gamma^*=\left(
\begin{array}{cccc}
0 & -1 & e^{i\pi/3} & e^{-i\pi/3} \\
-1 & 0 & e^{-i\pi/3} & e^{i\pi/3} \\
e^{i\pi/3} & e^{-i\pi/3} & 0 & -1 \\
e^{-i\pi/3} & e^{i\pi/3} & -1 & 0 \\
\end{array}
\right)
\end{gather}

We define two quantities for later convenience.
\bea
C&=&e^{i\phi_{12}}+e^{i\phi_{23}}+e^{i\phi_{31}}+\text{h.c.}=-3
\nonumber\\
R&=&\zeta_{12}\zeta_{13}^*+\zeta_{23}\zeta_{21}^*+\zeta_{31}\zeta_{32}^*+\text{h.c.}=-3
\label{eq:C,R}
\eea
where $\phi_{ij}=\phi_i-\phi_j$ and $C$ is evaluated for [111]-direction.

\label{sec: 4th perturbation}
\section{II. Lowest corrections Eqs. (\ref{eq:110}), (\ref{eq:111}) in the coupling constants $g_p$}
In this section, we derive Eqs. (\ref{eq:110}), (\ref{eq:111}) and investigate the relevance of additional exchanges Eq. (\ref{eq:OtherEx}). The assignment is to evaluate the coefficient $g_p$ of the ring exchange in the effective action Eq. (\ref{eq:3rd pert}). It involves the classical spin ice with alternative spin configurations $\uparrow\downarrow\uparrow\downarrow\uparrow\downarrow$ or $\downarrow\uparrow\downarrow\uparrow\downarrow\uparrow$ around a plaquette. \cite{hermele2004pyrochlore} Integrating out the virtual states imposed by Eqs. (\ref{eq:Ham}) and (\ref{eq:PolarZeeman}), it flips all 6 spins on the plaquette.

Given the perturbation $H_1=H_{\text{pseu}}^{\pm}+H_{\text{Z}}^{\pm}$ bringing about the spinon hopping, the coupling constant $g_p$ arised from 
\bea
&-&g_p\text{cos}(\triangledown\times A_{\textbf{r}\textbf{r}'})\vert_{\hexagon}
\nonumber\\
&=& \langle \overline{\hexagon} \vert P\Big[H_1(G_oH_1)^2+H_1(G_oH_1)^3 + \cdots \Big]P\vert \hexagon \rangle \quad \quad
\label{eq:pert}
\eea
where $\vert \hexagon \rangle$ and $\vert \overline{\hexagon} \rangle$ are flippable and resonance partners each other. We denote the non-interacting Green function $G_0=(1-P)(E-H_{\text{pseu}}^z-H_{\text{Z}}^z)^{-1}(1-P)\simeq (1-P)(-H_{\text{pseu}}^z)^{-1}(1-P)+O(h^2)$ with the projector operator $P$ onto the ground state manifold. 

\begin{figure}[b]
\subfloat[]{\label{fig:3rd}\includegraphics[width=0.46\textwidth]{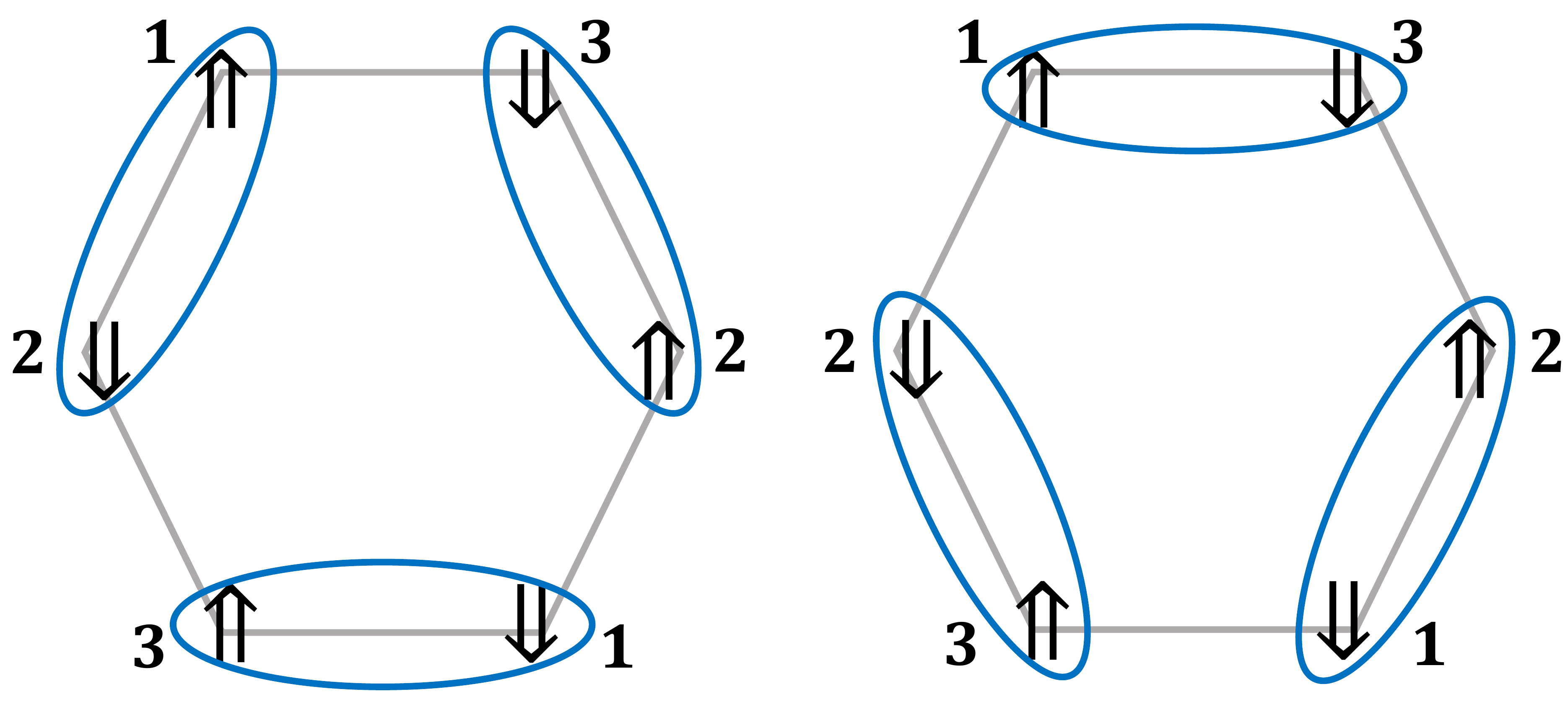}}
\nonumber\\
\subfloat[]{\label{fig:4thM1}\includegraphics[width=0.22\textwidth]{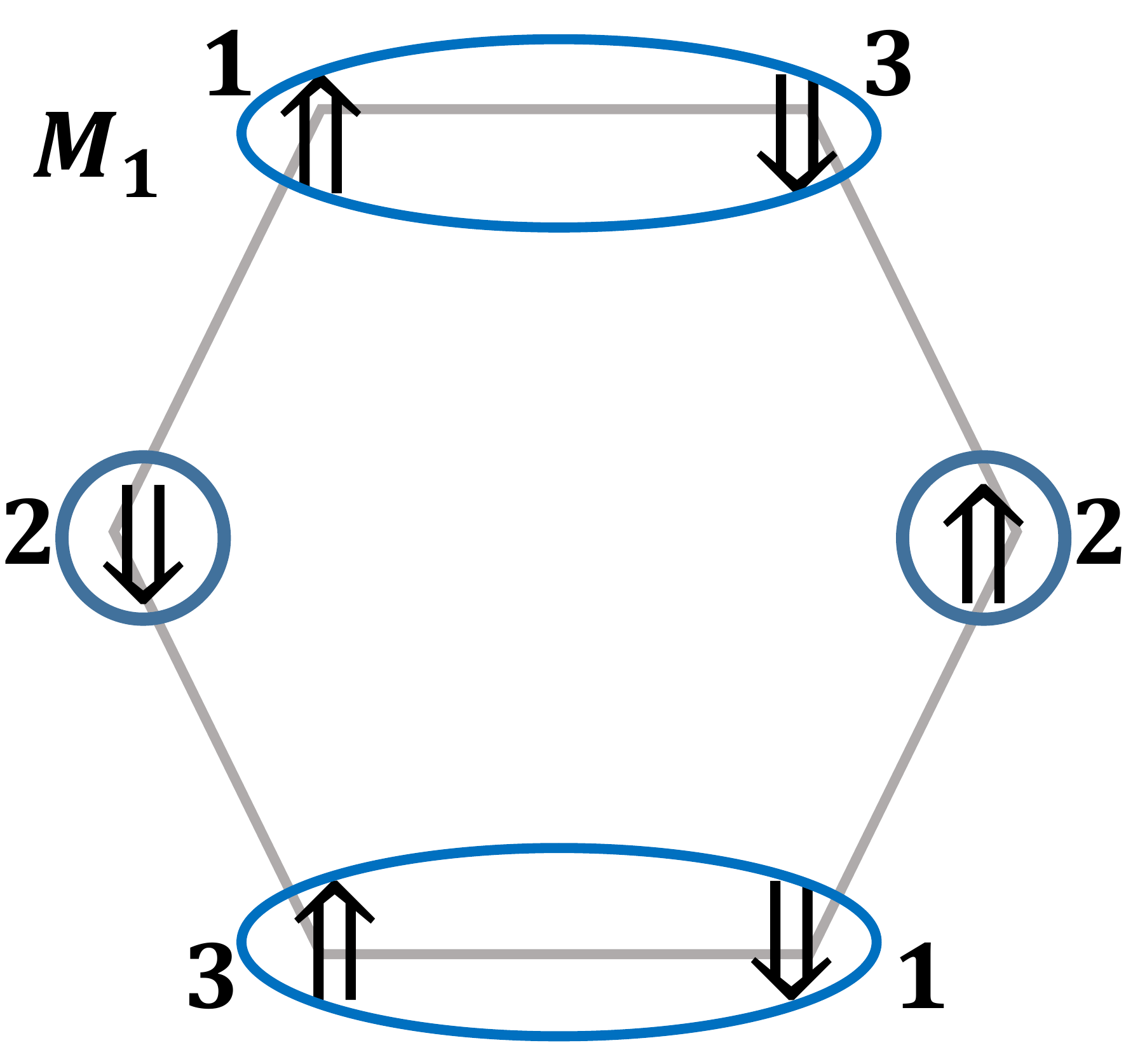}}
\quad
\subfloat[]{\label{fig:4thM2}\includegraphics[width=0.22\textwidth]{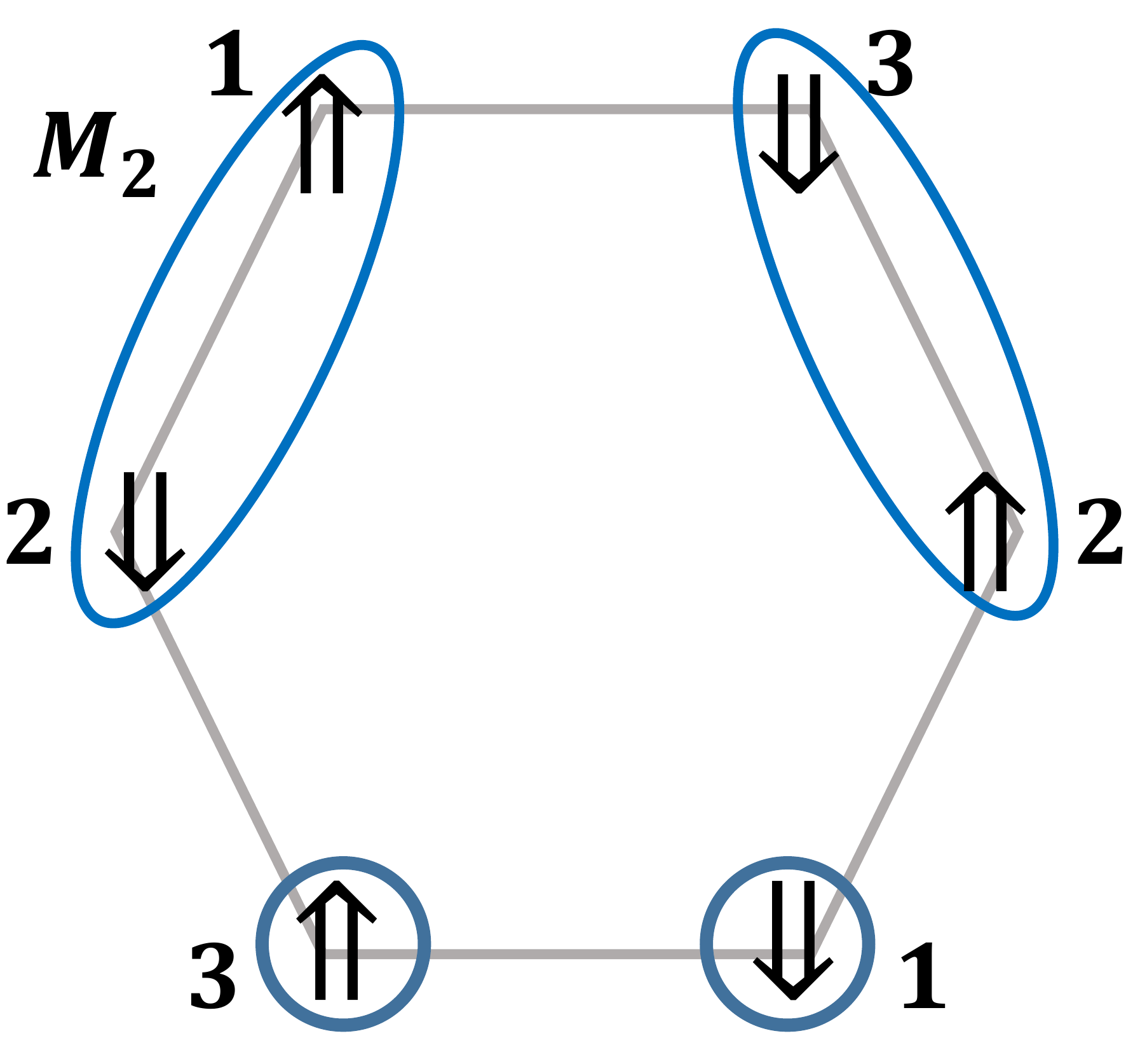}}
\caption{Abbreviated ring exchange process on the plaquettte containing the sublattice sites $i=1,2,3$. The exchange interaction $J_{\pm}S_i^+S_j^-$ and Zeeman coupling $h_i^{\perp}S_i^{\pm}$ are represented as the ellipse and unbiased circle respectively. Same circles are exchangeable in a sequence, which results in the ordering number $O$. The 3rd order perturbation Eq. (\ref{eq:3rdcoeff}) is contributed from (a) two different process with topologically equivalent configurations $M=2$, $O=3!$. Meanwhile, the 4th order perturbation is guided by two topologically different configurations, (b) $M_1$: the phases in Eq. (\ref{eq:PolarZeeman}) is canceled (c) $M_2$: not canceled  leaving an extra phase $\phi_{ij}$}
\label{fig:Process}
\end{figure}

The ring exchange Eq. (\ref{eq:pert}) is constructed from the alternative spin flip operators $S_1^-S_2^+S_3^-S_4^+S_5^-S_6^+$. The first term is the 3rd order correction constructed from the Eq. (\ref{eq:Ham}) only. The constant of proportionality counts the multiplicity of the topologically equivalent configuration $M$ and the order of intermediate states $O$. (Fig. (\ref{fig:3rd})) We are careful of the denominator since the propagators differently evolve on each sequence.
{For convenience, we abbreviate the intermediate process as the coefficients of interactions in Eqs. (\ref{eq:Ham}), (\ref{eq:PolarZeeman}) with a sequence from right to left.} 
For example, the multiplicity of the process $(-\frac{J_{\pm}}{2})(-\frac{J_{\pm}}{2})(-\frac{J_{\pm}}{2})$ is $M=2$ and the number of ordering is $O=3!$. Also all intermediate states contain 2 gauge charges $H_{\text{pseu}}^z=J_z$, the first term in Eq. (\ref{eq:pert}) yields
\bea
\text{3rd}: 2\times 3! \times \frac{(-J_{\pm}/2)^3}{J_z^2}=-\frac{3J_{\pm}^3}{2J_z^2}
\label{eq:3rdcoeff}
\eea

The lowest order where the Zeeman coupling Eq. (\ref{eq:PolarZeeman}) engages in is the second term in Eq. (\ref{eq:pert}), the 4th order correction $\sim J_{\pm}^2h^2/J_z^3$. Obviously, this depends on the external field direction and the relative angle between the field and the plaqueete. We start the enumeration with [110]-direction. Among 4 plaquettes, 2 plaquettes containing sublattice sites $i=0,1,3$ and $i=0,2,3$ are parallel to the B-field and the others with $i=0,1,2$ and $i=1,2,3$ are oblique. There are two distinct configurations $M_1=3, M_2=6$ and number of ordering is $O_1=O_2=2!\cdot 2!$ to be treated separately. (Fig. (\ref{fig:4thM1}), (\ref{fig:4thM2}))

On the plaquette with $i=0,1,3$, the process $(\frac{h_i^{\perp}}{2})(\frac{h_i^{\perp}}{2})(\frac{J_{\pm}}{2})(\frac{J_{\pm}}{2})$ for $M_1$ contributes 
\bea
\Big[(2!\cdot 2!) \times \Big(2(\frac{h^{\perp}_0}{2})^2+(\frac{h^{\perp}_1}{2})^2\Big)\Big]\times \frac{(J_{\pm}/2)^2}{-J_z\cdot 2J_z \cdot J_z}\quad\quad
\label{eq:[110]hhJJM1}
\eea
where each factor in the square bracket counts $O_1=4$ and $M_1=3$ keeping the dependence of $h_i^{\perp}$ on $i$. In the denominator, $2J_z$ takes into account the intermediate state with 4 charges. Likewise the process $(\frac{h_i^{\perp}}{2})(\frac{h_i^{\perp}}{2})(\frac{J_{\pm}}{2})(\frac{J_{\pm}}{2})$ for $M_2$ 
\bea
\Big[(&2!&\cdot 2!) \times  \Big(e^{i\phi_{03}}(\frac{h^{\perp}_0}{2})^2+(e^{i\phi_{01}}+e^{i\phi_{13}})(\frac{h^{\perp}_1}{2})^2+\text{h.c.}\Big)\Big]
\nonumber\\
&\times &\frac{(J_{\pm}/
2)^2}{-J_z\cdot J_z \cdot J_z}\quad\quad
\label{eq:[110]hhJJM2}
\eea
Also the process $(\frac{h_i^{\perp}}{2})(\frac{J_{\pm}}{2})(\frac{J_{\pm}}{2})(\frac{h_i^{\perp}}{2})$ for $M_1$ and $M_2$ are respectively
\bea
\Big[(&2!&\cdot 2!) \times \Big(2(\frac{h^{\perp}_0}{2})^2+(\frac{h^{\perp}_1}{2})^2\Big)\Big]\times \frac{(J_{\pm}/2)^2}{-J_z\cdot J_z \cdot J_z}\quad\quad
\label{eq:[110]hJJh M1}
\\
\Big[&2!& \times  \Big(e^{i\phi_{03}}(\frac{h^{\perp}_0}{2})^2+(e^{i\phi_{01}}+e^{i\phi_{13}})(\frac{h^{\perp}_1}{2})^2+\text{h.c.}\Big)\Big]
\nonumber\\
&\times & \Big(\frac{(J_{\pm}/
2)^2}{-J_z\cdot J_z \cdot J_z} + \frac{(J_{\pm}/
2)^2}{-J_z\cdot 2J_z \cdot J_z}\Big)
\label{eq:[110]hJJh M2}
\eea

There are 4 more sequences $(\frac{J_{\pm}}{2})(\frac{J_{\pm}}{2})(\frac{h_i^{\perp}}{2})(\frac{h_i^{\perp}}{2})$, $(\frac{h_i^{\perp}}{2})(\frac{J_{\pm}}{2})(\frac{h_i^{\perp}}{2})(\frac{J_{\pm}}{2})$ etc. Among them, the contribution $(\frac{J_{\pm}}{2})(\frac{J_{\pm}}{2})(\frac{h_i^{\perp}}{2})(\frac{h_i^{\perp}}{2})$ is same as $(\frac{h_i^{\perp}}{2})(\frac{h_i^{\perp}}{2})(\frac{J_{\pm}}{2})(\frac{J_{\pm}}{2})$ and the others are same as 
$(\frac{h_i^{\perp}}{2})(\frac{J_{\pm}}{2})(\frac{J_{\pm}}{2})(\frac{h_i^{\perp}}{2})$. Collecting them all, the 4th order correction on the plaquette $i=0,1,3$ is
\bea
&2&\times\Big(\frac{(J_{\pm}/2)^2}{-J_z^3}\Big\lbrace-(h_0^{\perp})^2+\frac{(h_1^{\perp})^2}{2}\Big\rbrace\Big)
\nonumber\\
&+&4\times\Big(\frac{(J_{\pm}/2)^2}{-J_z^3} \Big\lbrace\frac{(h_0^{\perp})^2}{2}+(h_1^{\perp})^2\Big\rbrace\Big)
=-\frac{5J_{\pm}^2h^2}{4J_z^3}\quad\quad
\label{eq:[110]013}
\eea
On the plaquette with $i=1,2,3$, all enumeration procedure is same accompanied by the interchange  $\phi_0 \leftrightarrow \phi_1$, $\phi_2 \leftrightarrow \phi_3$ and $h_{0,3}^{\perp} \leftrightarrow h_{1,2}^{\perp}$. Using Eq. (\ref{eq:[110]013}) on the plaquette $i=1,2,3$,
\bea
&2&\times\Big(\frac{(J_{\pm}/2)^2}{-J_z^3}\Big\lbrace-(h_1^{\perp})^2+\frac{(h_0^{\perp})^2}{2}\Big\rbrace\Big)
\nonumber\\
&+&4\times\Big(\frac{(J_{\pm}/2)^2}{-J_z^3} \Big\lbrace\frac{(h_1^{\perp})^2}{2}+(h_0^{\perp})^2\Big\rbrace\Big)
=-\frac{J_{\pm}^2h^2}{J_z^3}\quad\quad
\label{eq:[110]123}
\eea
Together with Eqs. (\ref{eq:3rdcoeff}), (\ref{eq:[110]013}) and (\ref{eq:[110]123}), we obtain Eq. (\ref{eq:110}).

Next, we work out the case with [111]-direction. On the perpendicular plaquette with $i=1,2,3$, the calculation is same as Eqs. (\ref{eq:[110]hhJJM1})-(\ref{eq:[110]hJJh M2}) with the substitution $h_0^{\perp} \rightarrow h_1^{\perp}$. The process $(\frac{h_i^{\perp}}{2})(\frac{h_i^{\perp}}{2})(\frac{J_{\pm}}{2})(\frac{J_{\pm}}{2})$ for the configurations $M_1$, $M_2$ are 
\bea
\Big[&(&2!\cdot 2!) \times 3\times (\frac{h^{\perp}_1}{2})^2\Big]\times \frac{(J_{\pm}/2)^2}{-J_z\cdot 2J_z \cdot J_z}\quad\quad
\label{eq:[111]on hhJJM1}
\\
\Big[&(&2!\cdot 2!) \times C \times (\frac{h_1^{\perp}}{2})^2 \Big] \times \frac{(J_{\pm}/2)^2}{-J_z\cdot J_z \cdot J_z}\quad\quad
\label{eq:[111]on hhJJM2}
\eea
Similarly, for the process $(\frac{h_i^{\perp}}{2})(\frac{J_{\pm}}{2})(\frac{J_{\pm}}{2})(\frac{h_i^{\perp}}{2})$, Eqs. (\ref{eq:[110]hJJh M1}) and (\ref{eq:[110]hJJh M2}) become
\bea
\Big[&(2!&\cdot 2!) \times 3\times (\frac{h^{\perp}_1}{2})^2\Big]\times \frac{(J_{\pm}/2)^2}{-J_z\cdot J_z \cdot J_z}\quad\quad
\label{eq:[111]on hJJhM1}
\\
\Big[&2!& \times C\times (\frac{h^{\perp}_1}{2})^2\Big]
\nonumber\\
&&\times
\Big(\frac{(J_{\pm}/2)^2}{-J_z\cdot J_z \cdot J_z} + \frac{(J_{\pm}/
2)^2}{-J_z\cdot 2J_z \cdot J_z}\Big)
\label{eq:[111]on hJJhM2}
\eea
Adding Eqs. (\ref{eq:[111]on hhJJM1})-(\ref{eq:[111]on hJJhM2}) altogher,
\bea
&&2\times\Big(\frac{3}{2}\frac{(J_{\pm})^2(h_1^{\perp}/2)^2}{-J_z^3}+C\frac{(J_{\pm})^2(h_1^{\perp}/2)^2}{-J_z^3} \Big)
\nonumber\\
&+&4\times\Big(3\frac{(J_{\pm})^2(h_1^{\perp}/2)^2}{-J_z^3}+\frac{3}{4}C\frac{(J_{\pm})^2(h_1^{\perp}/2)^2}{-J_z^3} \Big)=0\quad\quad\quad
\label{eq:[111]on}
\eea
On the other 3 tilted plaquettes, the multiplicity of $M_1$ is reduced from 3 to 2 since the B-field to parallel to the sublattice $i=0$, $h_0^{\perp}=0$. Similarly for $M_2$, no flipping occur on the site $i=0$ by the B-field, $C \rightarrow e^{i\phi_{12}}+\text{h.c.}=-1$. Thus Eqs. (\ref{eq:[111]on hhJJM1})-(\ref{eq:[111]on hJJhM2}) become
\bea
\Big[&(2!\cdot 2!)& \times 2\times (\frac{h^{\perp}_1}{2})^2\Big]\times \frac{(J_{\pm}/2)^2}{-J_z\cdot 2J_z \cdot J_z}\quad\quad
\label{eq:[111]out hhJJM1}
\\
\Big[&(2!\cdot 2!)& \times (-1) \times (\frac{h_1^{\perp}}{2})^2 \Big] \times \frac{(J_{\pm}/2)^2}{-J_z\cdot J_z \cdot J_z}\quad\quad
\label{eq:[111]out hhJJM2}
\\
\Big[&(2!\cdot 2!)& \times 2\times (\frac{h^{\perp}_1}{2})^2\Big]\times \frac{(J_{\pm}/2)^2}{-J_z\cdot J_z \cdot J_z}\quad\quad
\label{eq:[111]out hJJhM1}
\\
\Big[&(2!\cdot 2!)& \times (-1)\times (\frac{h^{\perp}_1}{2})^2\Big]
\nonumber\\
&&\times
\Big(\frac{(J_{\pm}/2)^2}{-J_z\cdot J_z \cdot J_z} + \frac{(J_{\pm}/
2)^2}{-J_z\cdot 2J_z \cdot J_z}\Big)
\label{eq:[111]out hJJhM2}
\eea
Likewise, the correction on the tilted plaquettes is
\bea
&2&\times\Big(\frac{J_{\pm}^2(h_1^{\perp}/2)^2}{-J_z^3}+\frac{J_{\pm}^2(h_1^{\perp}/2)^2}{-J_z^3}(-1)\Big)+4\times
\nonumber\\
\Big(&2&\frac{J_{\pm}^2(h_1^{\perp}/2)^2}{-J_z^3} +\frac{3}{4}\frac{J_{\pm}^2(h_1^{\perp}/2)^2}{-J_z^3}(-1) \Big)=-\frac{10}{9}\frac{J_{\pm}^2h^2}{J_z^3}\quad\quad\quad
\label{eq:[111]out}
\eea
Combining Eqs. (\ref{eq:3rdcoeff}), (\ref{eq:[111]on}) and (\ref{eq:[111]out}), we have Eq. (\ref{eq:111}).

Finally, we evaluate the additional exchange interactions Eq. (\ref{eq:OtherEx}). To embody the exchange $\sim S_i^{\pm}S_j^{\pm}$ into the ring exchange $S_1^-S_2^+S_3^-S_4^+S_5^-S_6^+$, we need to flip at least one site three times subsequently e.g. the process $(J_{\pm}S_1^-S_2^+)(J_{\pm}S_3^-S_4^+)(h_6^{\perp}S_6^+)(J_{\pm\pm}S_5^-S_6^-)(h_6^{\perp}S_6^+)$. Thus the lowest correction where this term participates is 5th order and will be disregarded in our discussion. Meanwhile, the interaction $\sim S_i^zS_j^{\pm}$ plays a role similar to the Zeeman term. 

First we consider the correction $\sim J_{\pm}^2J_{\pm z}h/J_z^3$. Since this term is odd in $h$ and $J_{\pm z}$, each intermediate process has its partner with reversed sequence, e.g. $(-\frac{J_{\pm}}{2})(-\frac{J_{\pm}}{2})(-\frac{h_i^{\perp}}{2})(\frac{J_{\pm z}}{2})$ and $(\frac{J_{\pm z}}{2})(-\frac{h_i^{\perp}}{2})(-\frac{J_{\pm}}{2})(-\frac{J_{\pm}}{2})$. Each pair shares the same $M, O$ except the sign $S_i^z$ merged in $J_{\pm z}S_i^zS_j^{\pm}$. This is because the sign $S_i^z$ relies on whether the exchange $S_i^{\pm}$ on $i$ happens before $S_i^zS_j^{\pm}$ or not. Thus all pairs cancel each other and the correction $\sim J_{\pm}^2J_{\pm z}h/J_z^3$ vanishes.

Now we consider the correction $\sim J_{\pm}^2J_{\pm z}^2/J_z^3$, which has nothing to do with the field direction and the plaquettes. 
The counting is similar to the Zeeman coupling except the additional factor $S_i^z$ acting together on the neighbor of $S_j^{\pm}$. Focusing on a single plaquette containing $i=1,2,3$, the multiplicity increases $2^2=4$ times to take account of the site where $S_i^z$ acts on. Similarly to Eqs. (\ref{eq:[110]hhJJM1}), (\ref{eq:[110]hhJJM2}), (\ref{eq:[111]on hhJJM1}), (\ref{eq:[111]on hhJJM2}), the process $(\frac{J_{\pm z}}{2})(\frac{J_{\pm z}}{2})(\frac{J_{\pm}}{2})(\frac{J_{\pm}}{2})$ for $M_1$ and $M_2$ are
\bea
4&\times& \Big[(-\zeta_{21}-\zeta_{23})(\zeta_{21}^*+\zeta_{23}^*)+(-\zeta_{13}-\zeta_{12})(\zeta_{13}^*+\zeta_{12}^*)
\nonumber\\
&+&(-\zeta_{31}-\zeta_{32})(\zeta_{31}^*+\zeta_{32}^*) \Big]\times \frac{(J_{\pm}/2)^2(J_{\pm z}/2)^2}{-J_z\cdot 2J_z \cdot J_z}
\label{eq:jjJJM1}
\\
&2&\times \Big[(\zeta_{32}^*-\zeta_{31}^*)(-\zeta_{12}-\zeta_{13})+(-\zeta_{12}+\zeta_{13})(\zeta_{31}^*+\zeta_{32}^*)
\nonumber\\ 
&+&1\rightarrow 3\rightarrow 2\rightarrow 1 + 1\rightarrow 2\rightarrow 3\rightarrow 1 + \text{h.c.}\Big]
\nonumber\\
&\times & \frac{(J_{\pm}/2)^2(J_{\pm z}/2)^2}{-J_z\cdot J_z \cdot J_z}\quad\quad
\label{eq:jjJJM2}
\eea
where the term with arrows is the duplicated one with migrated site index following the arrow. Each duplication counts the number of equivalent configurations 3 in Eq. (\ref{eq:[111]on hhJJM1}). From now on, these duplications will be omitted in $\dots$.
The complex numbers in the square bracket dwell on the sign and the site where $S_i^z$ acts on. Using the definition Eq. (\ref{eq:C,R}), these complex numbers are $-(6+R)$ and $2(6-R)$ respectevely.
The process $(\frac{J_{\pm}}{2})(\frac{J_{\pm}}{2})(\frac{J_{\pm z}}{2})(\frac{J_{\pm z}}{2})$ is same as Eqs. (\ref{eq:jjJJM1}), (\ref{eq:jjJJM2}).

The process $(\frac{J_{\pm z}}{2})(\frac{J_{\pm}}{2})(\frac{J_{\pm}}{2})(\frac{J_{\pm z}}{2})$ for the configuration $M_1$ and $M_2$
\bea
&4&\times \Big[(\zeta_{21}+\zeta_{23})(\zeta_{21}^*+\zeta_{23}^*)+\dots \Big]\times \frac{(J_{\pm}/2)^2(J_{\pm z}/2)^2}{-J_z\cdot J_z \cdot J_z}
\nonumber\\
&=&4(6+R)\frac{(J_{\pm}/2)^2(J_{\pm z}/2)^2}{-J_z\cdot J_z \cdot J_z}\quad\quad
\label{eq:jJJjM1}
\\
&2&\times\Big[(\zeta_{32}^*+\zeta_{31}^*)(\zeta_{12}+\zeta_{13}) +\text{h.c.}+\dots\Big]
\nonumber\\
&\times&\Big(\frac{(J_{\pm}/2)^2(J_{\pm z}/2)^2}{-J_z\cdot J_z \cdot J_z}
+\frac{(J_{\pm}/2)^2(J_{\pm z}/2)^2}{-J_z\cdot 2J_z \cdot J_z}\Big)
\nonumber\\
&=&2(6+3R)\frac{3}{2}\Big(\frac{(J_{\pm}/2)^2(J_{\pm z}/2)^2}{-J_z\cdot J_z \cdot J_z}\Big)
\label{eq:jJJjM1}
\eea
where the complex numbers in the bracket replace $3$ and $C$ in Eqs. (\ref{eq:[111]on hJJhM1}), (\ref{eq:[111]on hJJhM2}).

Likesie the process $(\frac{J_{\pm}}{2})(\frac{J_{\pm z}}{2})(\frac{J_{\pm z}}{2})(\frac{J_{\pm}}{2})$ for $M_1$ is
\bea
4&\times& \Big[(-\zeta_{21}+\zeta_{23})(-\zeta_{21}^*+\zeta_{23}^*)+\dots \Big]\times \frac{(J_{\pm}/2)^2(J_{\pm z}/2)^2}{-J_z\cdot J_z \cdot J_z}
\nonumber\\
&=&4(6-R)\frac{(J_{\pm}/2)^2(J_{\pm z}/2)^2}{-J_z\cdot J_z \cdot J_z}
\label{eq:JjjJM1}
\eea
and for $M_2$
\bea
\Big[(\zeta_{32}^*-\zeta_{31}^*)(-\zeta_{13}+&\zeta_{12}&)+(-\zeta_{12}+\zeta_{13})(\zeta_{31}^*-\zeta_{32}^*)+\dots \Big]
\nonumber\\
\times \frac{(J_{\pm}/2)^2(J_{\pm z}/2)^2}{-J_z\cdot J_z \cdot J_z}&=&2(6-R)\frac{(J_{\pm}/2)^2(J_{\pm z}/2)^2}{-J_z\cdot J_z \cdot J_z}\quad\quad\quad
\nonumber\\
\Big[(\zeta_{12}+\zeta_{13})(\zeta_{31}^*+&\zeta_{32}^*&)+(-\zeta_{31}^*-\zeta_{32}^*)(-\zeta_{12}-\zeta_{13})\dots \Big]
\nonumber\\
\times\frac{(J_{\pm}/2)^2(J_{\pm z}/2)^2}{-J_z\cdot 2J_z \cdot J_z}&=&2(6+3R)\frac{(J_{\pm}/2)^2(J_{\pm z}/2)^2}{-J_z\cdot 2J_z \cdot J_z}
\label{eq:JjjJM2}
\eea

The process $(\frac{J_{\pm z}}{2})(\frac{J_{\pm}}{2})(\frac{J_{\pm z}}{2})(\frac{J_{\pm}}{2})$ for $M_1$ vanishes since the process reverse the overall sign by interchanging two elements $(\frac{J_{\pm}}{2})$. For example, 
the process $(J_{\pm z}S_1^-S_2^z)(J_{\pm}S_2^+S_3^-)(J_{\pm\pm}S_4^+S_3^z)(J_{\pm}S_5^-S_6^+)$ cancel with the partner $(J_{\pm z}S_1^-S_2^z)(J_{\pm}S_5^+S_6^-)(J_{\pm\pm}S_4^+S_3^z)(J_{\pm}S_2^-S_3^+)$ due to $S_i^z$. For $M_2$ contribution,
\bea
\Big[(\zeta_{32}^*-\zeta_{31}^*)(-\zeta_{12}-&\zeta_{13}&)+(-\zeta_{12}+\zeta_{13})(\zeta_{32}^*+\zeta_{31}^*)+\dots \Big]
\nonumber\\
\times \frac{(J_{\pm}/2)^2(J_{\pm z}/2)^2}{-J_z\cdot J_z \cdot J_z}&=&2(6-R)\frac{(J_{\pm}/2)^2(J_{\pm z}/2)^2}{-J_z\cdot J_z \cdot J_z}\quad\quad\quad
\nonumber\\
\Big[(\zeta_{12}+\zeta_{13})(\zeta_{31}^*+&\zeta_{32}^*&)+(-\zeta_{31}^*-\zeta_{32}^*)(-\zeta_{12}-\zeta_{13})\dots \Big]
\nonumber\\
\times\frac{(J_{\pm}/2)^2(J_{\pm z}/2)^2}{-J_z\cdot 2J_z \cdot J_z}&=&2(6+3R)\frac{(J_{\pm}/2)^2(J_{\pm z}/2)^2}{-J_z\cdot 2J_z \cdot J_z}
\label{eq:jJjJM2}
\eea
where the bracket replaces $2!\times C$ in Eq. (\ref{eq:[111]on hJJhM2}). The process $(\frac{J_{\pm}}{2})(\frac{J_{\pm z}}{2})(\frac{J_{\pm}}{2})(\frac{J_{\pm z}}{2})$ is also same as Eq. (\ref{eq:jJjJM2}).

Gathering Eqs. (\ref{eq:jjJJM1})-(\ref{eq:jJjJM2}) altogether, we obtain the result $-9({J_{\pm}^2J_{\pm z}^2})/{J_z^3}$. In the presence of the exchanges such as Eq. (\ref{eq:OtherEx}), this correction is to be included in Eqs. (\ref{eq:110}) and (\ref{eq:111}), which decreases the critical fields.
\\
\\
\\
\\
\begin{figure}[!b]
\subfloat[]{\label{fig:Gaugefix1}\includegraphics[width=0.5\textwidth]{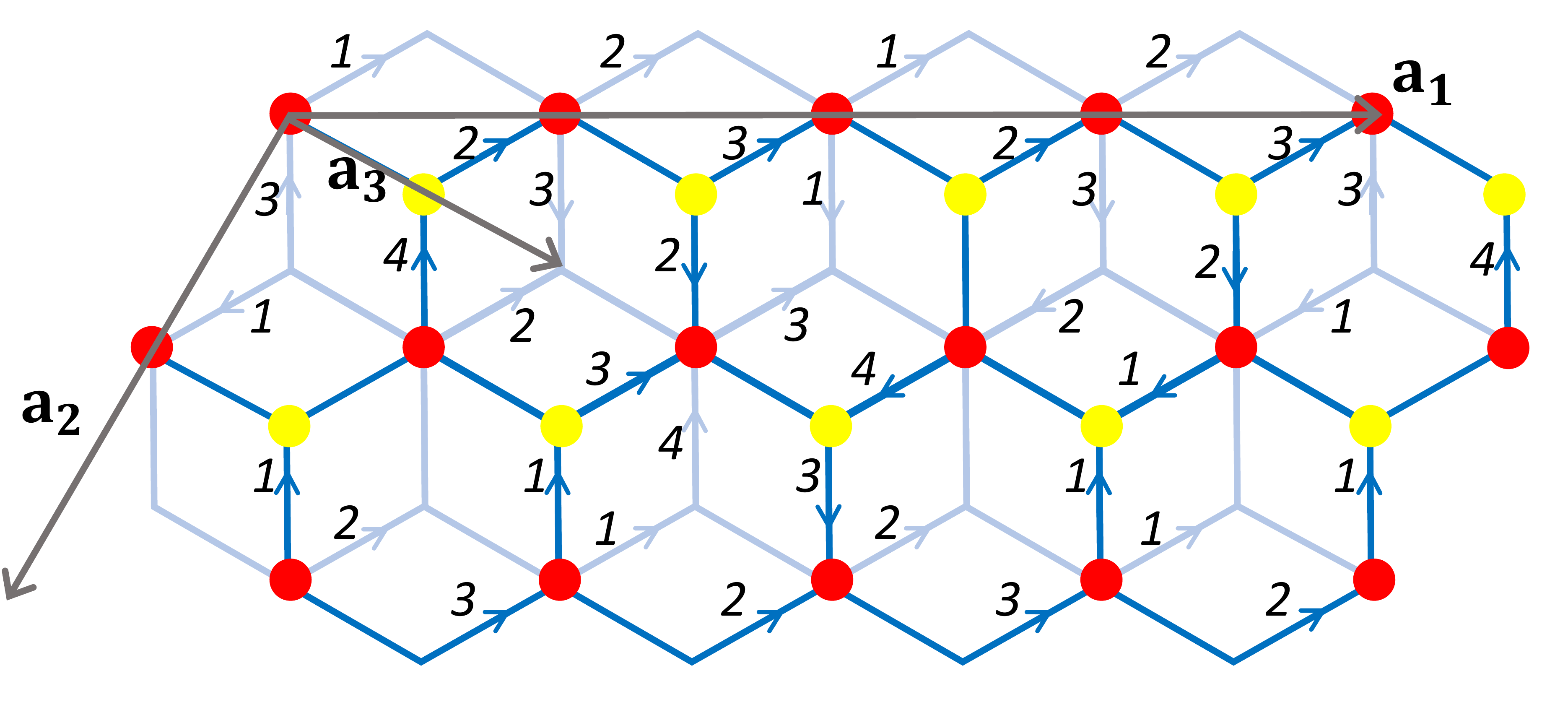}}
\\
\subfloat[]{\label{fig:Gaugefix2}\includegraphics[width=0.5\textwidth]{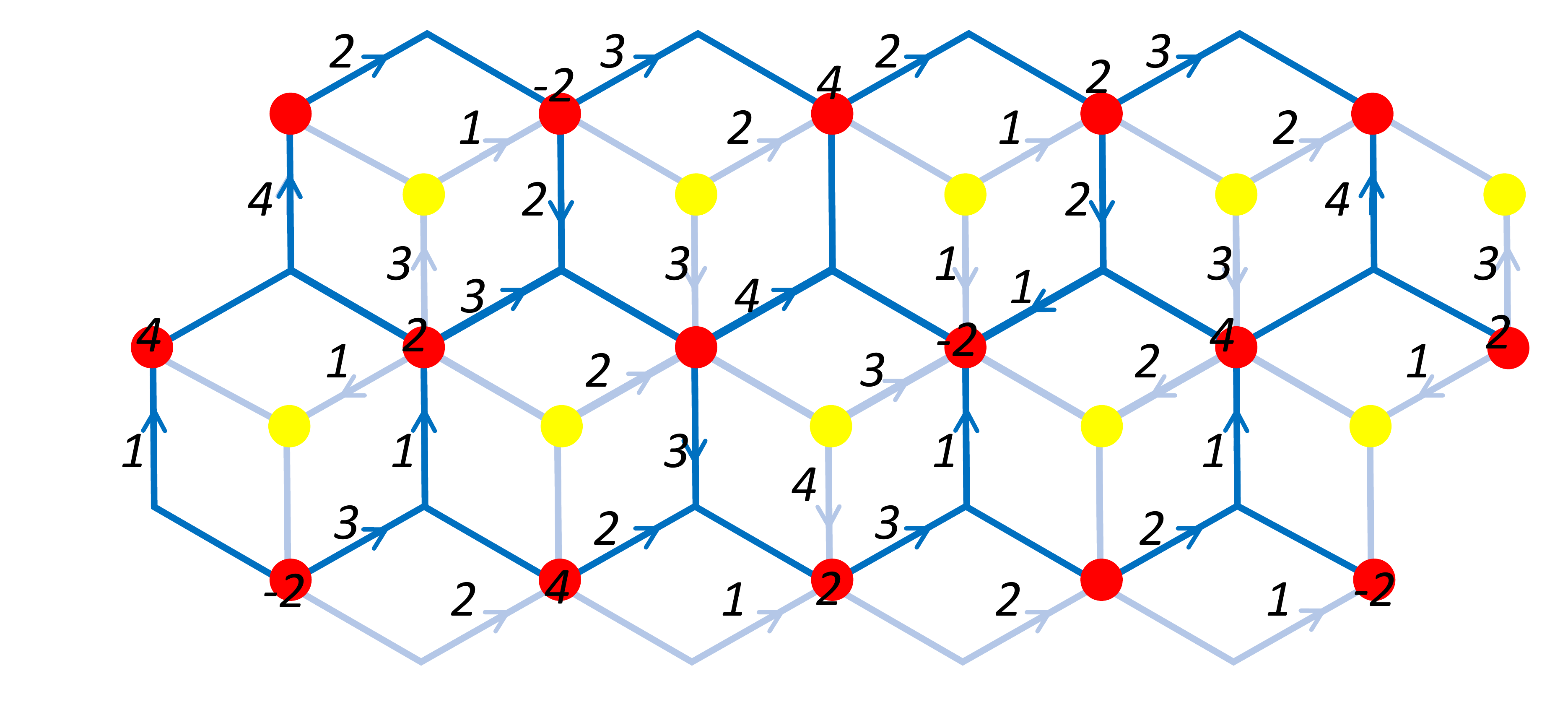}}
\caption{Gauge fixing with a flux $\mathcal{B}=\pi/4$ in Eq. (\ref{eq:Optimized B}) viewed from the B-field direction. 
The perpendicular and titled plaquettes trap the flux $3\mathcal{B}=3\pi/4$ and $\mathcal{B}=\pi/4$ respectively.
Each link is marked with the emergent gauge field $A_{ij}$ along the arrow in unit of $\pi/4$. The integers on the red sites mark the gauge field piercing out of the hexagonal plane. Since the magnetic translation vector $\textbf{a}_3$ is doublely extended with respect to the lattice constant, there are 2 hexagonal sheets in the magnetic unit cell. These views are
(a) between the height $z=0$ and $z=\hat{z}\cdot \textbf{a}_3/2$
and (b) between the height $z=\hat{z}\cdot \textbf{a}_3/2$ and $z=\hat{z}\cdot \textbf{a}_3$}
\label{fig:Gaugefix}
\end{figure}
\\

\label{sec: Gauge fixing}
\section{III. Gauge fixing on the diamond lattice with $\mathcal{B}=\pi/4$}

\begin{figure}[!h]
  \begin{center}
  \includegraphics[width=0.47\textwidth]{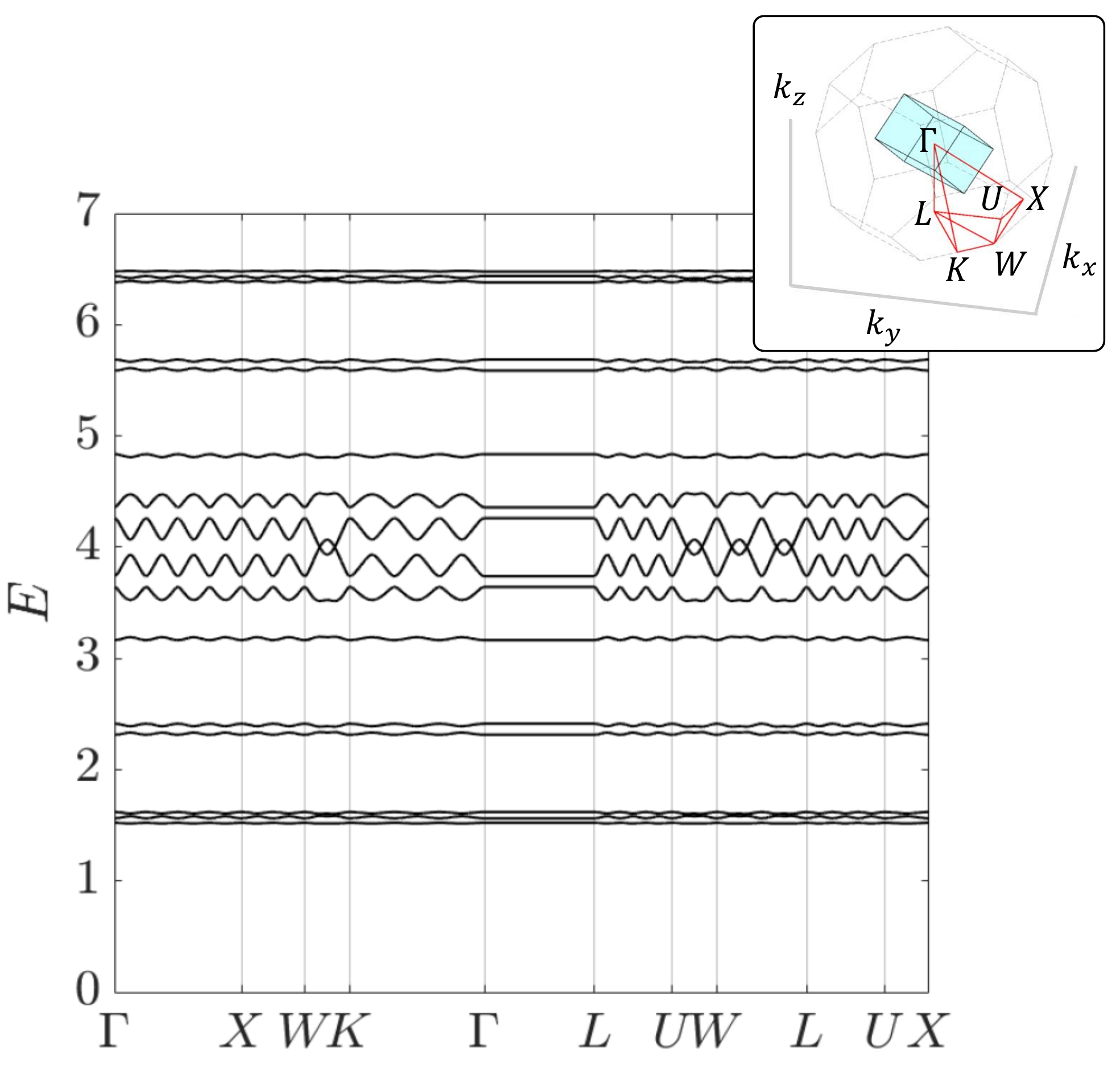}
\caption{
Spinon band spectrum along highly symmetric points in the Brillouin zone of the original fcc lattice. 
The dispersion energy is scaled in unit of the nearest neighbor hopping $t_{\textbf{r}\textbf{r}'}=1$.
(Inset) The Brillouin zone of the original fcc lattice and the reduced one. The reduced Brillouin zone is cuboid unlike the face-centered cubic Bravais lattice.
}
\label{fig:Spinonbands}
  \end{center}
\end{figure}

We now present the specific gauge choice to demonstrate the shaved flux $\mathcal{B}=\pi/4$ on the tilted plaquettes and $3\mathcal{B}=3\pi/4$ on the perpendicular plaquette in Eq. (\ref{eq:Optimized B}) when $\vert g\vert=\overline{g}$. In Fig. (\ref{fig:Gaugefix}), the emergent gauge field $A_{ij}$ in unit of $\pi/4$ is marked on the link connecting the diamond sites $i$ and $j$. The unit cell is enlarged $4\times 2\times 2=16$ times with the magnetic translation vectors $\textbf{a}_1=(4,0,0), \textbf{a}_2=(-1,\sqrt{3},0), \textbf{a}_3=(1,-1/\sqrt{3},2\sqrt{6}/3)$ in unit of the lattice constant $a$. Considering this gauge fixing, Fig.(\ref{fig:Spinonbands}) shows the spinon band structure keeping only the nearest neighbor hopping $t_{\textbf{r}\textbf{r}'}=1$ due to applying magnetic field and the chemical potential $\mu=-4$. 
The thermal hall coefficient Eq. (\ref{eq:ThermalHall}) is estimated based on this spinon band structure.  
We note that the spinon band structure and the behavior of the thermal hall conductivity could depend on the parameters ratios between $h/J_z, J_{\pm}/J_z$ and $J_{\pm z}/J_z$.
\\
\\
\\
\\
\\
\\
\\



\end{document}